\documentclass[usenatbib]{mn2e}

\usepackage{lscape,graphicx}

\newcommand{\microns}{$\mu$m}

\def\ga{\mathrel{\hbox{\rlap{\hbox{\lower4pt\hbox{$\sim$}}}\hbox{$>$}}}}
\def\la{\mathrel{\hbox{\rlap{\hbox{\lower4pt\hbox{$\sim$}}}\hbox{$<$}}}}
\def\msunyr{$M$\mbox{$_{\odot}$}\rm{yr}$^{-1}$}
\def\msun{$M$\mbox{$_{\normalsize\odot}$}}
\def\rsun{$R$\mbox{$_{\normalsize\odot}$}}
\def\mdot{$\dot{M}$}
\def\mstar{$M$\mbox{$_{\star}$}}
\def\mms{$M$\mbox{$_{\rm MS}$}}
\def\mfin{$M$\mbox{$_{\rm fin}$}}
\def\rgc{$R$\mbox{$_{\rm GC}$}}
\def\mcur{$M$\mbox{$_{\rm cur}$}}

\def\lsun{$L$\mbox{$_{\normalsize\odot}$}}
\def\lstar{$L$\mbox{$_{\star}$}}
\def\teff{$T$\mbox{$_{\rm eff}$}}
\def\lbol{$L$\mbox{$_{\rm bol}$}}
\def\tform{$t$\mbox{$_{\rm form}$}}
\def\Qlyc{$Q$\mbox{$_{\rm Lyc}$}}
\def\kms{\,km~s$^{-1}$}

\def\arcsec{$^{\prime \prime}$}

\def\sige{$\Sigma_{e^{-}}$}
\def\sigcl{$\Sigma_{\rm cl}$}
\def\ne{$n_{\rm e}$}
\def\hii{H{\sc ii}}
\def\RMS{{\it{RMS}}}
\def\MSX{{\it{MSX}}}
\def\SFR{{\it SFR}$_{\rm Gal}$}
\def\um{$\mu$m}
\def\tkh{$t_{\rm KH}$}
\def\chisq{$\chi^2$}
\newcommand{\fig}[1]{Fig.\ \ref{#1}}
\newcommand{\Fig}[1]{Figure \ref{#1}}
\newcommand{\eq}[1]{Eq.\ (\ref{#1})}

\title[Critical Tests of Accretion Models for Massive Stars]{The RMS
  Survey: Critical Tests of Accretion Models for the Formation of Massive Stars}
\author[B. Davies et al.]{Ben Davies$^{1,2}$, Melvin G.\ Hoare$^{2}$,
  Stuart L.\ Lumsden$^2$,  Takashi Hosokawa$^{5,6}$, \newauthor
  Ren\'{e} D.\ Oudmaijer$^2$, James S.\ Urquhart$^3$, Joseph C.\ Mottram$^{4}$ and
  Joseph Stead$^{2}$
 \\
$^{1}$Institute of Astronomy, University of Cambridge, Madingley Road,
 Cambridge CB3 0HA, UK\\
$^2$School of Physics \& Astronomy, University of Leeds, Woodhouse
  Lane, Leeds LS2 9JT, UK\\
$^3$Australia Telescope National Facility, CSIRO, Sydney, NSW 2052, Australia\\
$^4$School of Physics, University of Exeter, Exeter,
  Devon EX4 4QL, UK\\
$^5$Jet Propulsion Laboratory, California Institute of Technology,
  Pasadena CA 91109, USA\\
$^6$Department of Physics, Kyoto University, Kyoto 606-8502, Japan
}
\begin{document}

\date{Accepted ... Received ...}

\pagerange{\pageref{firstpage}--\pageref{lastpage}} \pubyear{2009}

\maketitle

\label{firstpage}

\begin{abstract}
There is currently no accepted theoretical framework for the formation
of the most massive stars, and the manner in which protostars continue
to accrete and grow in mass beyond $\sim$10\msun\ is still a
controversial topic. In this study we use several prescriptions of
stellar accretion and a description of the Galactic gas distribution
to simulate the luminosities and spatial distribution of massive
protostellar population of the Galaxy. We then compare the observables
of each simulation to the results of the {\it Red MSX Source (RMS)}
survey, a recently compiled database of massive young stellar
objects. We find that the observations are best matched by accretion
rates which increase as the protostar grows in mass, such as those
predicted by the {\it turbulent core} and {\it competitive accretion}
(i.e. Bondi-Hoyle) models. These `accelerating accretion' models
provide very good qualitative and quantitative fits to the data,
though we are unable to distinguish between these two models on our
simulations alone. We rule out models with accretion rates which are
constant with time, and those which are initially very high and which
fall away with time, as these produce results which are quantitatively
and/or qualitatively incompatible with the observations. To
simultaneously match the low- and high-luminosity YSO distribution we
require the inclusion of a `swollen-star' pre-main-sequence phase, the
length of which is well-described by the Kelvin-Helmholz
timescale. Our results suggest that the lifetime of the YSO phase is
$\sim 10^5$yrs, whereas the compact \hii-region phase lasts between
$\sim 2-4 \times 10^5$yrs depending on the final mass of the
star. Finally, the absolute numbers of YSOs are best matched by a
globally averaged star-formation rate for the Galaxy of 1.5-2\msunyr.




\end{abstract}


\section{Introduction}

The formation mechanism for stars with masses $\ga$10\msun\ is still
unclear. The steepness of the Initial Mass Function (IMF) means that
such stars are rare, while the short Kelvin-Helmholz contraction times
of massive stars mean that they arrive on the main-sequence (MS)
whilst still accreting and heavily embedded within their natal
molecular clouds. Meanwhile, from a theoretical point-of-view there is
considerable doubt that the `classical' theory of star formation is
applicable massive stars. Firstly, in the model of an isothermal
collapsing sphere, the accretion rate depends only on the sound-speed
of the gas \citep{Shu77}, and implies unrealistically large formation
timescales for stars with masses
$\ga$10\msun\ \citep{Stahler00}. Secondly, once the massive proto-star
arrives on the MS it will exert considerable outward radiation
pressure on the infalling material. In the spherically-symmetric case,
this can impede and eventually halt accretion, preventing the
formation of a high-mass star \citep{Kahn74,W-C87}. For further
discussions regarding both the theoretical and observational
challenges faced in the study of massive star formation, see
\citet{Z-Y07}.

There are several proposed amendments to the classical model of star
formation to account for massive stars, a few of which we now
discuss. \citet[][hereafter MT03]{M-T03} suggested a mechanism whereby
the larger molecular pre-stellar cores are supported against collapse
by larger turbulent motions (the {\it turbulent core} model). Once
these cores become unstable to collapse they have accretion rates
which accelerate with time. In the {\it competitive accretion} model
\citep{B-B06}, a large reservoir of gas is funnelled to the centre of
a cluster of protostars, allowing the cores at the cluster's centre to
achieve very large accretion rates through the Bondi-Hoyle
mechanism. \citet{S-K04} proposed that the accretion rates of
proto-stars within a gas cloud are governed by the complex interplay
of self-gravity and turbulence, and that upon collapse they have
initially very high accretion rates owing to the large amounts of
material that has accumulated in their immediate viscinity (the {\it
  gravo-turbulent} model). Indeed, several authors have proposed
models for star formation in which accretion rates behave in this way
with time
\citep{Foster-Chevalier93,Whitworth-WardThompson01,Motoyama-Yoshida03}.
Meanwhile, 3-D hydrodynamical simulations of massive forming stars
through accretion from a circumstellar disk displayed accretion rates
which were approximately constant throughout the simulation, though
they may pass through distinct accretion phases and be highly variable
on short timescales \citep{Krumholz09,Kuiper11}.

The mechanisms described above predict accretion rates which evolve
differently throughout the formation timescale of the star. The
turbulent core and competitive accretion (Bondi-Hoyle) models, though
they differ in form, both predict accretion rates \mdot\ that should
accelerate with time as the central star grows in mass. The
gravo-turbulent model however predicts that there should be an initial
burst of accretion, which gradually tails off as the star approaches
its final mass. These can be contrasted to the `standard model' of
star-formation, as well as those found in the numerical simulations of
\citet{Krumholz09} and \citet{Kuiper11}, when the accretion rate is
largely constant. As the accretion rate and its evolution will govern
the proto-star's luminosity as a function of time, as well as the
star's formation timescale, with a large complete sample of massive
forming stars it should be possible to distinguish between these
scenarios from analysis of their luminosity distribution.

\subsection{The \RMS\ Survey}

Such a sample has recently come to fruition. The {\it Red MSX Source
  (RMS)} survey \citep{Hoare05}\footnote{{\tt
    http://www.ast.leeds.ac.uk/RMS}} has selected point-sources from
the {MSX} catalogue \citep{Egan03} with colours appropriate for young
massive stars \citep{Lumsden02}, and weeded out contaminant objects
such as planetary nebulae and evolved stars through follow-up
near-infrared spectroscopy and continuum radio observations
\citep{Urquhart07a,Urquhart09a}. Sources have kinematic distance
estimates from molecular line observations
\citep{Urquhart07b,Urquhart08,Urquhart09b}, while the sources'
bolometric luminosities have been determined from fits to their
spectral energy distributions \citep{Mottram11}. High-resolution
mid-infrared imaging has been used to investigate source multiplicity
\citep{Mottram07}. The \RMS\ database now contains over 1,000
confirmed massive young stars. These objects are separated into
massive Young Stellar Objects (YSOs) and compact \hii-regions based on
one of two criteria. Objects which are spatially extended in the high
resolution mid-infrared images (spatial resolution $\sim$1\arcsec) due
to Lyman-$\alpha$ heating of the surrounding dust, or which have
detectable radio emission ($>$1mJy/beam) are considered to be ionizing
their surroundings, and hence are classified as \hii-regions. These
objects are considered to have reached the main-sequence, though they
may still be accreting. Objects displaying no evidence for an ionized
nebula are classified as YSOs. 

In this paper we use the results of the \RMS\ survey to test the
predictions of different accretion laws. We do this by constructing
realistic populations of massive protostars and distributing them
throughout the Galaxy according to a model of the Galaxy's gas
distribution. We compute the observed properties of the objects in the
simulation, apply the \RMS\ selection and classification criteria and
compare them to the observed results. 

This paper is similar in concept to that of \citet{Froebrich06}, who
used the numerical results of \citet{S-K04} to make predictions about
the earliest stages of star formation. It is also similar to
\citet{R-W10}, who calculated the evolution of the spectral energy
distribution (SED) of intermediate mass YSOs {\mstar$\sim$5-10\msun}
using a single accretion law. From this, they calculated the predicted
fluxes in the four Spitzer/IRAC channels and compared the total number
of objects to those found in the GLIMPSE survey. They then used this
information to predict the global star-formation rate of the
Galaxy. Our approach is different, in that we predict the
21\micron\ luminosity function of massive YSOs for a variety of
different accretion scenarios, with the ultimate goal of understanding
how massive stars form, although we too are able to put some
constraints on the Galactic global star-formation rate.

We begin in Sect.\ \ref{sec:model} with a description of the model
construction. The results of the model, the effect of the free
parameters, and of the different accretion laws, are presented in
Sect.\ \ref{sec:results}. We discuss the results and the implications
for models of massive star formation in Sect.\ \ref{sec:accrates}. We
provide a summary of our results in Sect.\ \ref{sec:conc}.


\section{Model construction} \label{sec:model}

The methodology for the construction of the model is as follows:
assuming a global star formation rate of the Galaxy, we generate
$10^{6}$yrs of stars -- up until the point where the oldest massive
stars have comfortably reached the main-sequence and have broken free
of their natal environments. Using a description of the gas
distribution in the Galaxy, these stars are each inserted into the
model Galaxy at a random location dictated by the relation between
local gas density and local star-formation rate. That is, locations of
high density have higher star formation rates, and hence are more
likely to host newly-formed stars. Quantitatively, we scale the
star-formation rate with gas density {\it SFR}$\propto \rho^{N}$, in
accordance with the Schmidt-Kennicutt law for star-forming galaxies
\citep{Kennicutt98}. Typically, the value of $N\approx1.4$ is quoted,
though some authors have suggested that $N$ is closer to unity, and
may even be as low as 0.8 \citep{Blanc09}. Strictly this law applies
to the {\it surface} density $\Sigma_{\rm dens}$ of galaxies, but
since the scale height of the Galaxy's gas distribution is small
compared to the radial size, $\rho \propto \Sigma_{\rm dens}$. 

Each star in the simulation is given an age (i.e.\ time since
accretion first began), which is randomly selected from a uniform
distribution between 0 -- $10^6$yrs. Using a given model of stellar
accretion, the age, and final mass \mfin\ of each star, we compute the
current mass \mcur. The pre-main-sequence birthlines of
\citet{H-O09} are then used to compute the luminosities of each star,
which we then use to calculate the observed flux at 21\micron. For
stars which we determine have arrived on the main-sequence, making
standard assumptions we calculate the properties of the surrounding
\hii-region, and ultimately the observed radio flux. We then pass the
observables through the \RMS\ selection criteria, and those sources
which would be detected in the survey are then classified according to
their observed properties. Below, each step of the process is
described in more detail.

\subsection{Galactic gas distribution} \label{sec:galcube}
For a prescription of the distribution of baryonic material in the
Galaxy, we use as a basis the Galactic distribution of free electrons
as determined by \citet[][ and refs therein]{T-C93}, and later updated
by \citet{C-L02} in a model known as NE2001. These studies use
pulsars with known distances to infer the line-of-sight density of
free electrons. We assume that in the Milky Way disk, the total ISM
density (neutral gas, dust etc.)  scales directly with the density of
free electrons.

\begin{figure}
  \centering
  \includegraphics[width=8.5cm,bb=20 0 566 552]{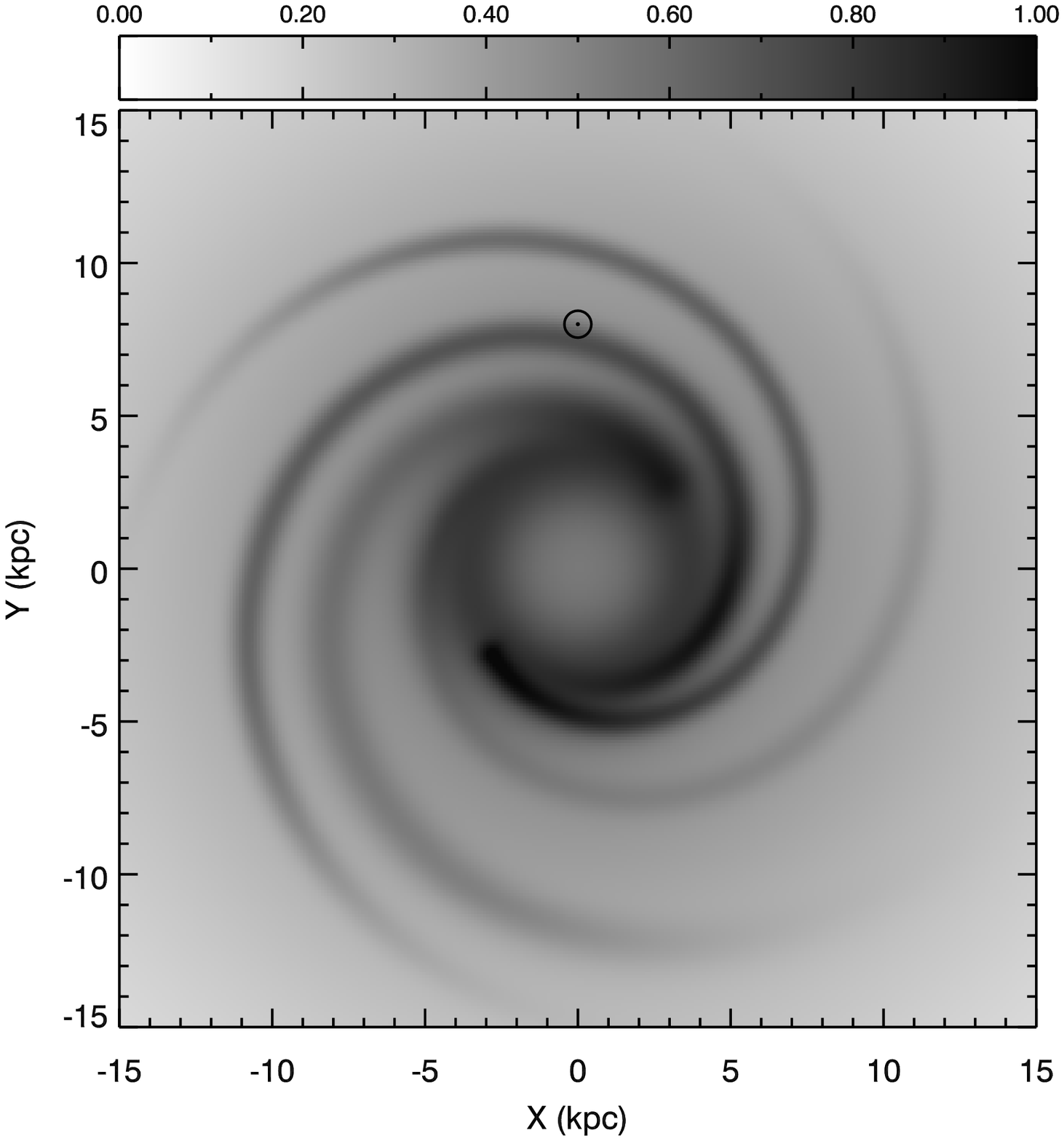}
  \includegraphics[width=8.5cm,bb=20 0 566 198]{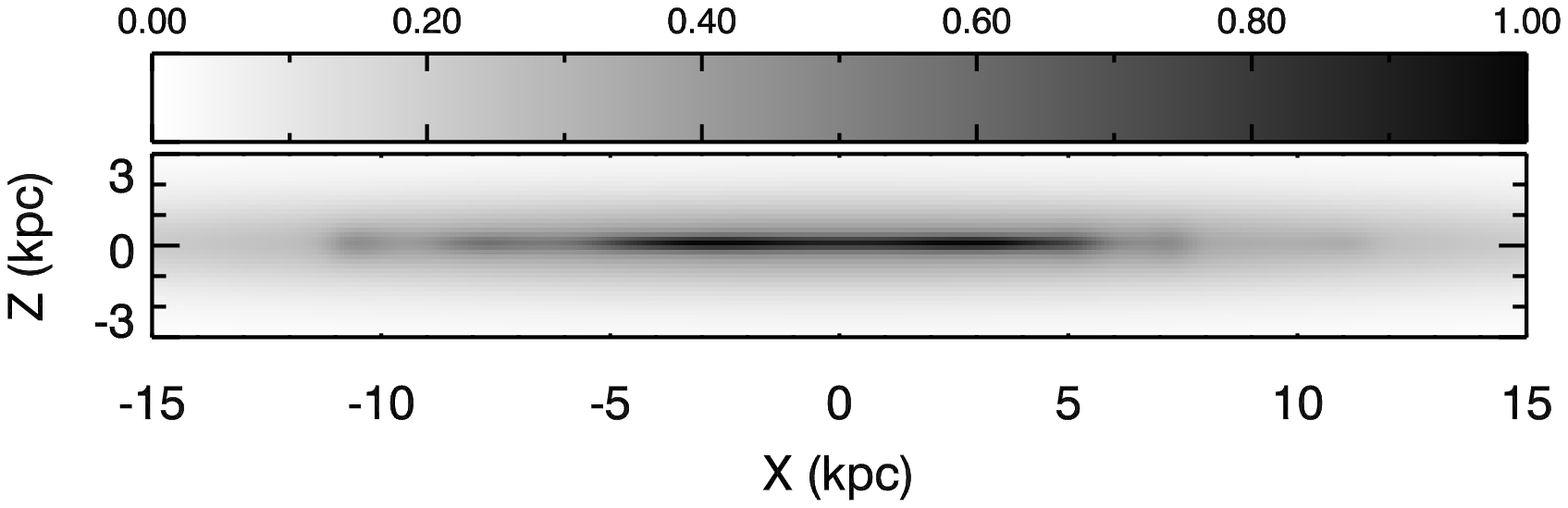}
  \caption{Gas surface densities of the model Galaxy in the $XY$ (top)
    and $XZ$ (bottom) planes. The panels show the linear grey-scale of
    the normalised gas density, with the relevant scale shown above
    each panel. The thick-disk, thin-disk and spiral-arm components,
    as described in TC93, are included. The location of the sun is
    indicated on the top panel.}
  \label{fig:galdens}
\end{figure}

The model consists of three components: thin disk, thick disk, and four
spiral arms. For the thin and thick disk components we use the
definitions and physical parameters as defined in NE2001. For the
spiral arms, we have made slight adjustments: we used the coordinates
of the fiducial points for each of the four arms as quoted by TC93 and
fitted each arm with a logarithmic spiral,

\begin{equation} 
\theta_{j} = A_{sp, j} \log(r_{j}/r_{min, j}) + \theta_{0, j} 
\label{equ:sp}
\end{equation}

\noindent where $\theta_{j}$ and $r_{j}$ define the Galactocentric
distance and azimuthal angle\footnote{$\theta_{j} \equiv 0$ is defined
  as the Sun - Galactic Centre axis.} of each spiral arm $j$; $A_{sp,
  j}$ and $\theta_{0, j}$ are fitted constants; and $r_{min, j}$ is
the minimum Galactocentric radius of the defining fiducial points in
TC93. We specified that two spiral arms should originate from the ends
of the central Bar. To ensure this we adjusted the fiducial points
such that $r_{min, j}$ was always 4kpc \citep[the length of the Bar as
  determined by ][]{Benjamin05} and $\theta_{0, j}$ was 45\degr\ for
the Scutum-Crux arm and 225\degr\ for the opposing (Perseus) arm. The
other two arms (Sagittarius and Norma) were traced inwards according
to the logarithmic fit, and the inner regions of the arms were
attenuated to blend with the inner disk component using an attenuation
function $S$, similar to that used in the NE2001 model;

\begin{eqnarray*}
  S = & {\rm sech}^{2}(r-A_{\rm inner}) & {\rm for~} (r-A_{\rm inner}) < 0 \\
  S = & 1                             & {\rm for~} (r-A_{\rm inner}) > 0
\end{eqnarray*}

\noindent We used a somewhat arbitrary value of 5kpc for $A_{\rm
  inner}$ in order to blend them with the inner thin disk. For the
outer disk, the spiral arms were extrapolated radially outwards to
21kpc, at which point their contribution to the local gas density is
negligible. The normalised surface density of the Galaxy gas
distribution model in both the $XY$ and $XZ$ planes can be seen in
Fig.\ \ref{fig:galdens}.

The description of the Galactic gas density we use has not been
precisely tuned to the current state-of-the-art picture of our
Galaxy's morphology. For example, we do not include the `kinks' in the
arms in the Solar neighbourhood, nor do we include other smaller
features such as the `local' arm. Our goal here is not to accurately
predict the spatial distribution of the YSO population, nor is it to
trace the spiral structure of our Galaxy. The purpose of this work is
to model the observed luminosity distributions of protostars, and so
we only need a description of the Galaxy's morphology which is
realistic in terms of size-scales (e.g. the scale-height, $z$) and
discrete structures (e.g. the number and size-scales of spiral
arms). Tweaks that we {\it could} make to the model gas distribution
to better match the spatial distribution will be discussed later,
while a better match to the 3-dimensional distribution of young
massive stars awaits the implementation of more accurate distances
\citep[see e.g.][]{Reid09}.

\begin{figure*}
  \centering
  \includegraphics[width=17cm]{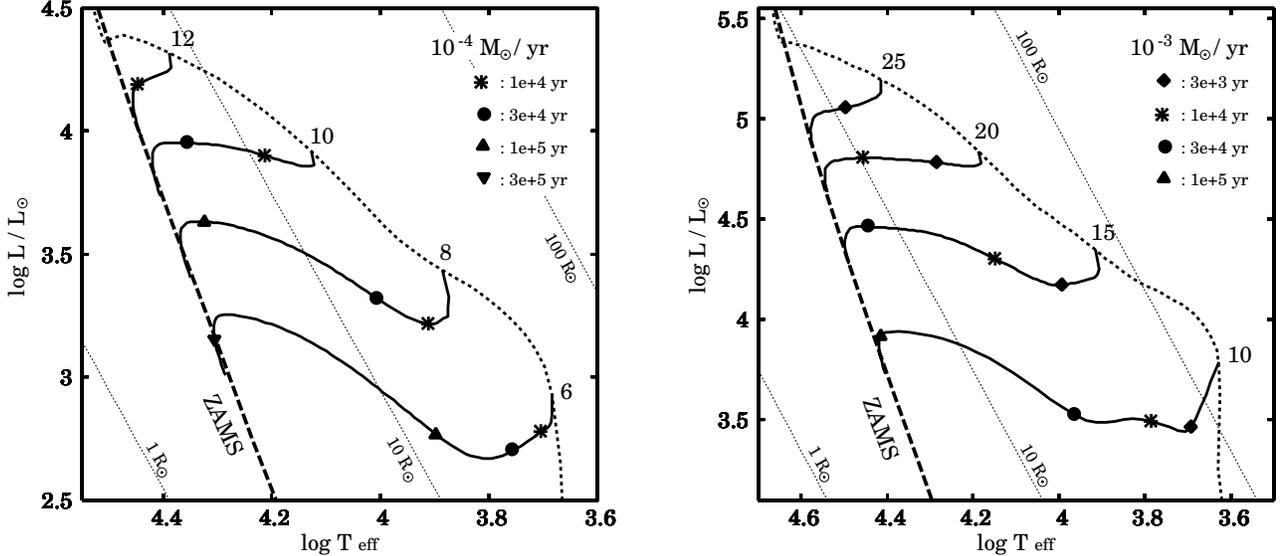}
  \caption{Evolution tracks of massive pre-main-sequence stars for the
    original accretion rates of $\dot{M} = 10^{-4}$\msunyr\ (upper
    panel) and $10^{-3}$\msunyr\ (lower panel) in the H-R diagram. The
    dotted line represents the birth line at these accretion rates in
    each panel. In the upper (lower) panel, the solid lines present
    the evolution tracks of 12, 10, 8, and 6\msun\ (25, 20, 15, and
    10\msun) stars after mass accretion ceases. The symbols on each
    track indicate the elapsed time since the termination of mass
    accretion, $10^4$, $3 \times 10^4$, $10^5$, and $3 \times 10^5$
    years. The dashed line presents locii of zero-age main-sequence
    stars taken from \citet{Schaller92}. Stellar radius is
    constant at $R_* =1~R_\odot$, $10~R_\odot$, and $100~R_\odot$
    along the dotted lines.  }
  \label{fig:hr_pms}
\end{figure*}

\subsection{Generation of stellar populations}
Assuming a global star-formation rate for the Galaxy of \SFR, we
generate 1Myr of star-formation, under the assumption that any star
older than 1Myr would be sufficiently free of its natal material to be
precluded by the \RMS\ selection criteria as being too
evolved. Estimates of \SFR\ range from $\sim$2-10\msunyr, though most
measurements appear to be in the range 3-4\msunyr\ \citep[see][ and
  references therein]{Diehl06}. Recently, in a study which was similar
to the one presented here, \citet{R-W10} attempted to derive
\SFR\ from a sample of YSOs selected from GLIMPSE, finding that values
of \SFR=0.7-1.5\msunyr\ were preferred. This is somewhat lower than
the other estimates described above. However, \citet{R-W10} used only
one description of accretion history in their study, that of
\citet{B-M96}, though it is unclear whether they used the constant
accretion rates or those that increased with protostellar mass
\citep[see][]{Robitaille06}. In either case, these accretion rates of
\citet{B-M96} are an order of magnitude lower than those presented in
more contemporary studies of e.g.\ MT03. This produces slightly longer
pre-MS lifetimes and therefore results in larger numbers of YSOs. This
may explain in part why \citet{R-W10} required values of \SFR\ which
were lower than in previous studies. For now we adopt
\SFR=3\msunyr. This parameter affects only the overall normalization
of objects, and the effect on our conclusions of allowing this
parameter to vary slightly will be explored later.

For the total mass of stars in our model, we generate a distribution
of stellar masses according to the Initial Mass Function (IMF) of
\citet{Kroupa01}, which is essentially the Salpeter law but augmented
at low masses. The masses we generate are the {\it final} stellar
masses \mfin, i.e. the mass that each star will have once it has
finished accreting. We make the assumption that the global star
formation rate has been constant over the last 10$^6$yrs, and the age
of each star in the simulation is selected from a uniform random
distribution of ages from zero to 10$^6$yrs. We did experiment with
`clustering' the star formation into localized bursts, which is likely
to be a more realistic representation of how star formation proceeds
within the Galaxy, but this was found to have no impact on the results
other than to increase stochastic effects within a single simulation.

\subsection{Computation of observables}
In order to calculate the observed properties of the protostars in our
simulations, we utilize the theoretical evolution tracks for accreting
protostars taken from \citet[][ hereafter HO09]{H-O09}. The authors
calculated the evolution of accreting protostars for various constant
accretion rates. They show that the evolution with high accretion
rates of $\geq 10^{-4}~M_\odot~{\rm yr}^{-1}$ has some characteristic
features. With $10^{-3}~M_\odot~{\rm yr}^{-1}$, for example, the
protostar becomes significantly swollen and does not reach the
main-sequence (MS) for $M_* < M_{\rm MS} \simeq 30M_\odot$. The
critical mass $M_{\rm MS}$ becomes higher with the higher accretion
rate. If the mass accretion ceases before the arrival to the MS, the
star experiences a contraction phase lasting approximately a
Kelvin-Helmholtz timescale $t_{\rm KH}$,

\begin{equation}
t_{\rm KH} = \frac{GM^{2}_{\star}} {R_{\star} L_{\star}}
\label{equ:kh}
\end{equation}

This predicts that massive pre-main-sequence (PMS) stars exceeding
$8~M_\odot$ should exist.  Figure \ref{fig:hr_pms} shows the
evolution tracks for such massive PMS stars in the HR diagram. In this
paper, we assume that the massive PMS stars contract keeping the
constant luminosity for simplifying the detailed evolution.

\begin{figure*}
  \centering
  \includegraphics[width=17cm]{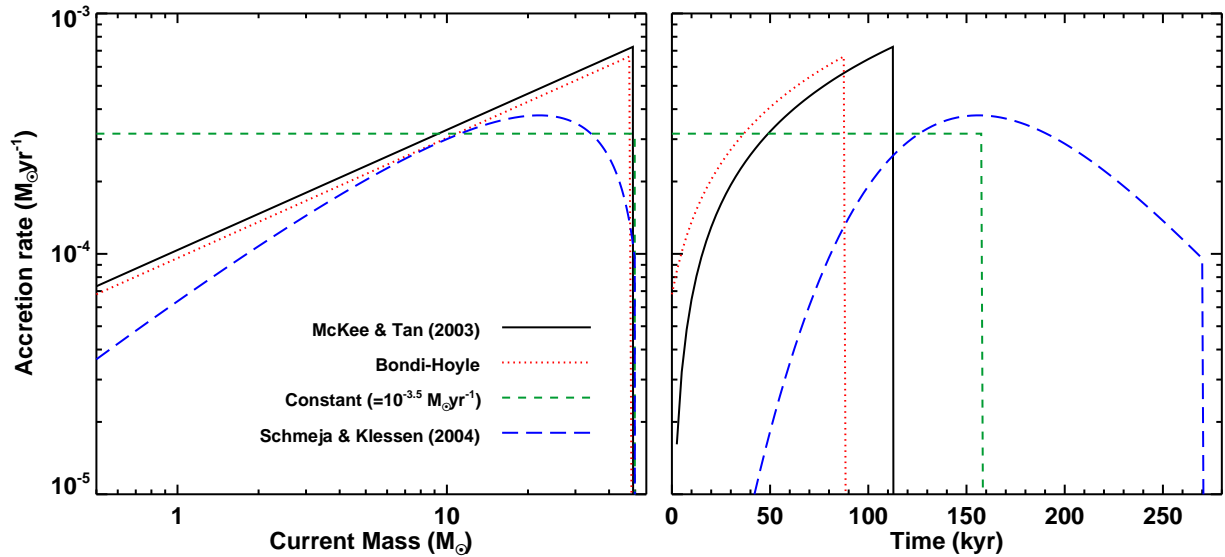}
  \caption{Evolution of the accretion rate for star with a final mass
    \mfin\ of 50\msun\ for the different prescriptions of accretion
    studied here. The panels show the accretion rate as a function of
    current mass \mcur\ (left) and as a function of time (right). The
    MT03 models plotted use \sigcl=1\,g\,cm$^2$. 
  }
  \label{fig:accrates}
\end{figure*}

\subsubsection{Accretion laws} \label{sec:acclaws}

To calculate the physical properties of the forming star, we must
first know how much matter has been accreted since the star began to
form. For this, we require a prescription of the star's accretion
rate. In this study we investigate three distinct accretion-rate
scenarios: constant accretion-rate; accretion-rates which increase
as the mass of the star grows; and those which decrease. Below, we describe the
prescriptions for each of these accretion rates, while in
\fig{fig:accrates} we illustrate the behaviour of the accretion rates
with both mass and time for each of these laws.

\paragraph*{a) Constant accretion}

Here we use the results of HO09 directly. For the age $t$ and final
mass \mfin\ of each star in our simulation, we calculate the current
mass for a specified accretion rate \mdot\ from,

\begin{equation}
M_{\rm cur} = \dot{M} t
\end{equation}

\noindent We can also calculate the formation timescale of the star
$t_{\rm form}$ from,

\begin{equation}
t_{\rm form} = M_{\rm fin} / \dot{M}
\end{equation}

We use the HO09 tracks to calculate the luminosity and temperature of
each star from its current mass \mcur. For stars which have reached
their final mass but whose masses are below the $M_{\rm MS}$ for that
accretion rate, we calculate the Kelvin-Helmholz timescale $t_{\rm
  KH}$. If a star is not older than $(t_{\rm form} + t_{\rm KH})$ then
it is considered to still be `swollen', that is it has not yet
contracted to the main-sequence, and so emits a negligible amount of
ionizing flux. Stars which {\it have} arrived on the main-sequence are
considered to have Lyman fluxes similar to zero-age main-sequence
stars of the same mass, but with extra luminosity due to accretion.

\paragraph*{b) Accelerating accretion I: MT03}
In the MT03 prescription of accretion by a massive protostar in the
turbulent core model, the accretion rate \mdot\ and formation
timescale \tform\ of an accreting star are given by
their Equs.\ (41) and (44),

\begin{eqnarray}
\dot{M} = 4.6 \times 10^{-4}M_\odot {\rm yr^{-1}}
\displaystyle\left(\frac{M_{\rm fin}}{30M_\odot}\right)^{0.75}
\Sigma_{\rm cl}^{0.75} \left(\frac{M_{\rm cur}}{M_{\rm
    fin}}\right)^{0.5}
\label{equ:mdot}
\end{eqnarray}

\begin{equation}
t_{\rm form} = 1.29\times10^{5} {\rm yr} 
\displaystyle\left(\frac{M_{\rm fin}}{30M_{\odot}}\right)^{0.25} 
\Sigma_{\rm cl}^{-0.75}
\label{equ:tform}
\end{equation}

\noindent where $\Sigma_{\rm cl}$ is the surface density of a
prestellar clump in units of g\,cm$^{-2}$. From these equations we can
find an expression for the current mass
$M_{\rm cur}$ of an accreting star,

\begin{eqnarray}
    M_{\rm cur} = & 0.18 M_{\odot} 
\displaystyle\left(\frac{M_{\rm fin}}{30M_{\odot}}\right)^{0.5}
 \Sigma_{\rm cl}^{1.5} ~ \left( \frac{t}{\rm 10^4 yr}\right)^{2} 
  \label{equ:mcur}
\end{eqnarray}

\noindent Using these equations, and the randomly generated stellar
ages, we can then calculate the current mass and formation timescale
of each star. 

To calculate the luminosities of the accreting star, we must take into
account the stellar luminosity and the accretion luminosity. To do
this we take the star's current mass and current accretion rate, and
look up the corresponding luminosity from the tracks of HO09. In doing
so, we are making two approximations: firstly, we linearly interpolate
the HO09 tracks in log-space at 0.01dex increments between 10$^{-5}$
and 10$^{-3}$\msunyr, since HO09 only calculate tracks at 10$^{-3}$,
10$^{-4}$ and 10$^{-5}$\msunyr. Secondly, we are assuming that a star
whose accretion rate has accelerated to the current value has a
similar luminosity to a star whose accretion rate has always been
constant at the current value. While one may expect there to be
systematic differences between these two cases, we found that the
\mcur-\lbol\ relationship of our approximation was quantitatively
similar to that computed by MT03 (their Fig.\ 6), since the accretion
luminosity only contributes a small fraction to the total luminosity.



\noindent For stars which have finished accreting, luminosties are
taken from the stellar structure models of \citet{Mey-Mae00}. 

Again, MT03 find that very massive stars join the MS at $M_{\rm MS}
\approx$20\msun, though the precise value of $M_{\rm MS}$ depends on
\sigcl. As with the case of constant accretion, stars which have
finished accreting but have final masses below $M_{\rm MS}$ are
assumed to undergo a pre-MS contraction phase which approximates to
$t_{\rm KH}$ for that particular star. 

\paragraph*{c) Accelerating accretion II: Bondi-Hoyle accretion}

Similar accelerating accretion is predicted under the Bondi-Hoyle (BH)
accretion relevant to the {\it competitive accretion} model
\citep{B-B06}. The difference here is that the accretion rate depends
only on $t$ and \mcur, not on \mfin. The BH accretion rate is given
by Equ.\ (6) of \citet{B-B06},

\begin{equation}
\dot{M} = 4 \pi \rho \frac{(GM_{\rm cur})^2}
{v_{i}^{3} (M_{\rm cur}/M_{i})^{3/2} }
\label{equ:bh_formal}
\end{equation}

\noindent where $v_{i}$ and $\rho$ are the velocity dispersion and gas
density within the star-forming clump, and $M_{i}$ is the initial mass
of a star-forming clump that becomes unstable to gravitational
collapse. Integrating and rearranging this equation, and using the
values determined by \citet{B-B06} for the parameters of $v_{i}$,
$M_{i}$ and $\rho$ (0.5\kms, 10$^{-17}$g\,cm$^3$, and
0.5\msun\ respectively), the current mass of a star accreting in this
way can be shown to be given by,

\begin{equation}
M_{\rm cur} = 0.5M_{\odot} ~ ( t / 10^4 {\rm yr} + 1 )^2
\label{equ:bh}
\end{equation}

\noindent The formation timescale is then simply,

\begin{equation}
t_{\rm form} = 10^4{\rm yr} ~ \displaystyle \left( \sqrt{ \frac{2M_{\rm
        cur}}{M_{\odot}} } - 1 \right)
\label{equ:bh}
\end{equation}

\noindent There has yet to be a thorough computation of BH birthlines,
so we must make assumptions for the object's luminosity, the duration
of the pre-MS `swollen' phase and the mass at which very massive stars
arrive on the MS, \mms. The luminosity of a BH accreting object is
again assumed to be well represented by the HO09 track with the same
current accretion rate at the same \mcur. We again assume that the MS
contraction time is equal to \tkh, and we leave \mms\ as a free
parameter in this model.

\paragraph*{d) Decelerating accretion}
For a representitive model of decelerating accretion, we use the
`gravo-turbulent fragmentation' model of \citet[][ hereafter
  SK04]{S-K04}. In this model, protostars at the centre of a cluster's
potential well are subject to a greater amount of infalling material,
and the initial accretion rates are very high. As the reservoir of
material is depleted, the accretion-rate gradually decreases,
eventually falling to zero. Before preceeding with a description of
how we incorporate this accretion scenario into our simulations, we
first note that in SK04 no stars more massive that 10\msun\ are
formed, and that these authors' goals were not necessarily to simulate
massive star formation. Nevertheless, we include it in our study as a
tool with which to examine how decelerating accretion affects the
relative numbers of massive protostars. 

For a prescription of such accretion, we use the recipes of SK04,
which are taken from empirical fits to their numerical
simulations. Here the accretion rate is given by,

\begin{equation}
\log(\dot{M}) = (\log \dot{M}_{0}) \frac{e}{\tau} t e^{-t/\tau}
\label{equ:sk04}
\end{equation}

\noindent where \mdot$_{0}$ and $\tau$ are constants. The constant
\mdot$_{0}$ specifies the {\it maximum} accretion rate, which is
reached after a brief period of rapid acceleration in \mdot. Once the
accretion rate reaches \mdot$_{0}$, it falls away gradually on a
timescale dictated by $\tau$.

As a representative case of decelerating accretion, we use the
simulation `G2' from SK04, which uses a gaussian realization
of the initial turbulence within the molecular cloud. Here, the value
of $\log$\mdot$_{0}$ was found to be a linear function of
$\log$\mfin, up to \mfin$\approx$10\msun. Specifically, from fits to
their data, 

\begin{equation}
\log(\dot{M_0}) = 0.75 \log M_{\rm fin} - 4.7
\label{equ:sk04_mdot}
\end{equation}

\noindent Though SK04 formed no stars with masses greater than
10\msun, for our simulations we make the simple assumption that this
relation can be extrapolated up to the most massive stars in our
simulation, \mfin=150\msun. The parameter $\tau$ also appears to be a
linear function of $\log$\mfin, though there is considerable
dispersion in this relation. From our linear fits to the results of
SK04, we find,

\begin{equation}
\tau / 10^{4}{\rm yrs} = \tau_{1} \log M_{\rm fin} + \tau_{0}
\label{equ:sk04_tau}
\end{equation}

\noindent with,
\begin{eqnarray*}
\tau_{0} & = &  (4.6 \pm 0.3) \times 10^{4}{\rm yrs}\\
\tau_{1} & = &  (4.3 \pm 2.0) \times 10^{4}{\rm yrs}
\end{eqnarray*}

\noindent Therefore, for a star of given \mfin\ and age $t$,
Eqs.\ (\ref{equ:sk04}), (\ref{equ:sk04_mdot}) and (\ref{equ:sk04_tau})
can be used to calculate that star's accretion rate as a function of
time since $t$=0. Integration of this function then yields the star's
current mass \mcur.

Using these parameters however, the formation times of massive stars
become unrealisticly long, and no stars with \mfin$\ga$15\msun\ can be
formed within 10$^{6}$yrs. Massive stars {\it can} be formed within a
reasonable time, but only if the constants in
Eq.\ (\ref{equ:sk04_tau}) are pushed to their extreme limits, such
that the largest possible values of $\tau$ result. This will be
discussed further in Sect.\ \ref{sec:accrates}.

As with the other accretion laws studied, luminosities are determined
from the pre-MS tracks of HO09, \mms\ is left as a free parameter, and
the length of the pre-MS `swollen' phase is assumed to be given by the
KH timescale.

\subsubsection{21\micron\ luminosity}

Having calculated \lbol\ from the accretion law, we can now determine
the observed flux through the {\it MSX} 21\microns\ filter, one
criterion used for source selection in the RMS survey. In
\citet{Mottram11} it was found that the 21\micron\ flux $F_{\rm 21}$
of objects in the sample scaled as a constant fraction of the total
bolometric flux $F_{\rm tot}$ according to,

\begin{equation}
 \log \displaystyle\left( \frac{F_{\rm tot}}{F_{\rm 21}} \right) = 
1.43 \pm 0.27
\label{equ:f21rat}
\end{equation}

\noindent No significant correlation was found between $F_{\rm
  tot}/F_{\rm 21} (\equiv C_{\rm 21})$ and any observable parameter in
the \RMS\ survey, such as IR colour or $F_{\rm 21}$.

To convert our $L_{\rm bol}$
determined in in Sect.\ \ref{sec:acclaws} to $F_{\rm 21}$, we use the
following relation,

\begin{equation}
F_{\rm 21} = \frac{L_{\rm bol}}{4 \pi D_{\odot}^{2}} 
\frac{\Delta_{21}}{10^{C_{21}}} 10^{(-A_{21}/2.5)}
\label{equ:f21}
\end{equation}

\noindent where $C_{21}$ is randomly drawn from a normal distribution
of numbers with a mean of 1.43 and standard deviation of 0.27
(following Eq.\ \ref{equ:f21rat}); $A_{\rm 21}$ is the interstellar
extinction (see below), and $\Delta_{21}$ is the effective bandwidth
of the {\it MSX} 21\micron\ filter.

As yet we have not dealt exlicitly with the issue of line-of-sight
extinction. There are likely two sources of extinction to each source
-- interstellar, and circumstellar. Circumstellar extinction will vary
largely from source to source, since it will depend on random
properties such as cavity opening angle and inclination to
line-of-sight. We have assumed that the factor $C_{\rm 21}$
incorporates the attenuating effect of circumstellar extinction.

In terms of interstellar extinction at 21\microns\ towards each
object, $A_{\rm 21}$, one may expect a systematic trend of this factor
with object distance. We have made a detailed study of this factor and
how it may impact our results, using Galactic star clusters as test
points (see Appendix A). Though we do account for this effect by
incorporating a factor of $10^{(-A_{21}/2.5)}$ into
Eq.\ \ref{equ:f21}, we expect at most that this plays a very minor
role in the results of our simulation, since $A_{\rm 21} \la 1.5$ for
even the most distant objects.

The current model does not account for a mid-IR faint or `dark' phase
early on in the evolution of massive stars. Examples of luminous young
objects that are mid-IR faint/dark clearly do
exist. \citet{Ellingsen06} found that 30\% of methanol maser sources,
that are invariably associated with massive YSOs, do not have
counterparts in the 4-8 micron GLIMPSE survey. Also, about 20\% of the
active cores in infrared dark clouds (IRDCs) have a luminosity
greater than 5000\lsun\ \citep{Rathborne10}. Though we are not yet
explicitly accounting for an IR `dark' phase here, during the very
early stages of the growth of massive stars their luminosity will be
below 5000\lsun\ and therefore not in the RMS survey; an aspect that
is already accounted for in the modelling presented here. Since very
few IR-dark YSOs have luminosities that would impact on our analysis,
we neglect such objects in this current work.


\begin{table*}
  \caption{Table of ZAMS stellar parameters as a function of mass. The
    relationship between stellar mass, temperature, luminosity and
    radius is taken from \citet{Mey-Mae00}. The photometric properties
    BC$_V$ and M$_V$ and the Lyman photon flux $Q_{\rm Lyc}$ as a
    function of \teff\ are taken from \citet{Martins-Plez06} and
    \citet{L-H07}, and are corrected for differences in
    \lstar. $^\dagger$These values are likely to be slightly
    underestimated, see text for details.  } \centering
  \begin{tabular}{lccccccc}
    \hline
    Mass   & T$_{\rm eff}$ & Log (L$_{\star}$/\lsun) & R$_{\star}$ &
    BC$_{V}$ & M$_{V}$ & Log$Q_{\rm Lyc}$   & t$_{\rm KH}$ \\
(\msun)& (K)           &               & (\rsun)     &          &         & (phot\, s$^{-1}$) & ($\times$10$^3$yrs) \\ 
\hline
\hline
  6 & 19000 & 3.01 &  3.0 & -1.83 &  -0.95 & 43.33 &  367.1 \\
  9 & 22895 & 3.57 &  3.9 & -2.28 &  -1.89 & 44.76 &  175.5 \\
 12 & 26743 & 3.96 &  4.5 & -2.64 &  -2.52 & 46.02 &  107.8 \\
 15 & 29783 & 4.26 &  5.1 & -2.88 &  -3.01 & 47.03 &   76.2 \\
 20 & 33824 & 4.61 &  6.0 & -3.22 &  -3.56 & 48.00 &   51.8 \\
 25 & 36831 & 4.86 &  6.7 & -3.47 &  -3.93 & 48.46 &   40.2 \\
 30 & 38670 & 5.02 &  7.3 & -3.61 &  -4.19 & 48.69 &   36.7 \\
 35 & 40596 & 5.18 &  8.0 & -3.75 &  -4.45 & 48.90 &   31.7 \\
 40 & 42610 & 5.34 &  8.7 & -3.89 &  -4.71 & 49.09 &   26.3 \\
 50 & 44589 & 5.52 &  9.8 & -4.03 &  -5.02 & 49.31 &   24.1 \\
 60 & 46647 & 5.70 & 11.0 & -4.18 &  -5.33 & 49.43 &   20.4 \\
 70 & 47662 & 5.81 & 12.0 & -4.25 &  -5.53 & 49.51$^\dagger$ &   19.7 \\
 80 & 48694 & 5.92 & 13.1 & -4.32 &  -5.74 & 49.58$^\dagger$ &   18.2 \\
 90 & 49449 & 6.02 & 14.1 & -4.37 &  -5.92 & 49.65$^\dagger$ &   17.4 \\
100 & 49913 & 6.09 & 15.0 & -4.41 &  -6.06 & 49.70$^\dagger$ &   17.0 \\
110 & 50381 & 6.16 & 16.0 & -4.44 &  -6.21 & 49.76$^\dagger$ &   16.4 \\
120 & 50853 & 6.23 & 17.1 & -4.47 &  -6.36 & 49.81$^\dagger$ &   15.5 \\
\hline
  \end{tabular}
  \label{tab:logq}
\end{table*}

\begin{figure}
  \centering
  \includegraphics[width=8.5cm]{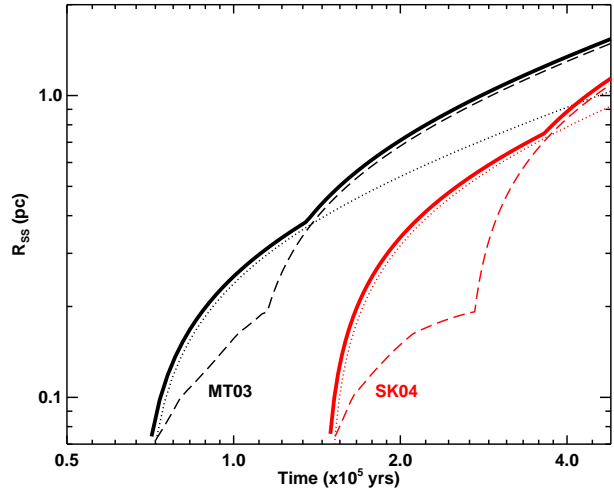}
  \caption{The evolution of Stromgren radius $R_{\rm SS}$ of a
    \hii-region around an accreting protostar, for two of the four different
    accretion scenarios studied here (with \mfin=50\msun). The dotted
    lines illustrate the pressure driven expansion of the \hii-region
    once the star hits \mms (=20\msun), under the false assumption
    that the ionizing flux remains constant thereafter. The dashed
    lines show the growth of $R_{\rm SS}$ in two different regimes:
    firstly, expansion at constant density due to the increase in
    ionizing photons as the star gains mass; and secondly,
    pressure-driven expansion once the mass has reached \mfin. The two
    approximations give different \hii-region sizes, and therefore
    give different radio fluxes. We make the simplifying assumption
    that the \hii-region size at any time is the larger of these two
    descriptions, illustrated by the solid line (offset for clarity). }
  \label{fig:rstrom}
\end{figure}

\subsubsection{Lyman flux}
\label{sec:lyman}

In the \RMS\ survey, massive YSOs are distinguished from young stars
which have recently arrived on the MS by assuming that the latter
class of objects are ionizing their surroundings and driving a
\hii-region. That is, sources with detectable \hii-regions are at a
more advanced evolutionary state than YSOs. Therefore, in our model we
must determine the ionizing power of a star that is identified as
having arrived on the MS. We then calculate the size and
radio-brightness of the \hii-region, and then whether or not the
\hii-region would be detected by the \RMS\ survey. If the source has
radio surface brightness above 1mJy/beam at 5GHz, then we assume that
it would be detected by the \RMS\ continuum radio observations
\citep{Urquhart07a,Urquhart09a}. If the \hii-region has a size larger
than 2\arcsec\ in diameter, the source would be identified as being
spatially extended in the high-resolution mid-infrared \RMS\ imaging
\citep{Mottram07}. If the source does not exceed these limits, it is
classified as a YSO.

In calculating these quantities, we begin with a star with mass
$M_{\rm cur}$ and determine the zero-age MS (ZAMS) stellar temperature
and luminosity from \citet{Mey-Mae00} at this mass. We then use the
calculations of \citet{Martins05} and \citet{L-H07}, which provide the
rate per unit area at which photons with frequencies above the Lyman
limit $Q_{\rm Lyc}$ are emitted for a star of a given temperature for
O and B stars respectively. The relationship between stellar
parameters and Lyman flux that we obtain is given in Table
\ref{tab:logq}.

There are several important points that should be noted regarding the
numbers in Table \ref{tab:logq}. Firstly, the calibrations of
\citet{Martins05} and \citet{L-H07} do not provide us with Lyman
fluxes for stars with masses above $\sim$60\msun. Since the
relationship between Lyman flux per unit area ($q_{\rm Lyc}$) and
stellar mass flattens off at high masses, we make the assumption that
$q_{\rm Lyc}$ is constant above 60\msun. This will mean that we are
likely to be slightly underestimating the Lyman flux of very massive
stars. However, this will only impact the number of \hii-regions at
high luminosities ($\ga 10^{5.7}$\lsun), which are very few in number
in the \RMS\ survey.

Secondly, the absolute Lyman photon rates per unit time $Q_{\rm Lyc}$
as a function of stellar mass that we compute are different to those
in \citet{Martins05} and \citet{L-H07}. This is because these authors
use a different calibration between stellar mass and temperature to
this study. Here, we use the relationship between \mstar,
\lstar\ \teff\ from the hydrostatic stellar models of
\citet{Mey-Mae00}.

Finally, we state explicitly that the final calibration of $Q_{\rm
  Lyc}$ versus $M_{\star}$ should be considered to be uncertain at the
30-50\% level, judging by the variation of numbers in the
literature. However, the effect of this uncertainty on our results
does not affect our conclusions, as will be shown later.

\subsubsection{Size and radio emission of \hii-regions}
\label{sec:radio}

Accurately calculating the radio emission from a young star is a
complex problem as there are many effects that one needs to consider:
the Lyman flux from the star, the density and morphology of the
ionized gas, the hydrodynamical expansion of the gas, and the
expansion of the ionization region due to the increasing Lyman flux
from the accreting star as it grows in mass. However, accurate radio
fluxes and expansion rates are not the goal of this project, more that
we want to get an estimate of the radio flux and the \hii-region size
in order to determine how the source would be classified in the
\RMS\ survey. As such, we are able to make a number of simplifying
assumptions when calculating the \hii-region preperties.

Firstly, when the star's ionizing flux is switched on, either at the
end of the `swollen star' phase, or as the mass of the star reaches \mms,
an \hii-region is formed with an initial size given by that of a
Str\"{o}mgren sphere,

\begin{equation}
R_{S} = \displaystyle\left( \frac{3Q_{\rm Lyc}}{4 \pi
  n_{\rm e}^{2}\beta_{\rm B}} \right)^{\frac{1}{3}}
\label{equ:rss0}
\end{equation}

\noindent where \ne\ is the initial electron density and $\beta_{\rm
  B} = 2.59 \times 10^{-13}{\rm cm^{3}s^{-1}}$ is the radiative
recombination coefficient of H$^{+}$ to all levels with $n \ge$2. We
then assume that one of two processes occurs: either the \hii-region
undergoes pressure-driven expansion; or that the \hii-region grows in
size at constant density due to the increase in Lyman flux from the
accreting star, followed by pressure-driven expansion once the star
reaches \mfin. We then determine which of these results in the largest
\hii-region. The differences between these two processes, in terms of
the size evolution of the \hii-region, is illustrated in
\fig{fig:rstrom}.

For pressure driven expansion of an \hii-region \citep[see
  e.g.][]{Dyson-Williams},

\begin{eqnarray}
R_{\rm S} & = & R_{\rm S,0} \displaystyle\left( 1 + 
\frac{7}{4}\frac{t}{t_{\rm cross}} \right)^{\frac{4}{7}} \\
n_{\rm e} & = & n_{\rm e, 0} \displaystyle\left( 
\frac{R_{\rm S,0}}{R_{\rm S}} \right)^{\frac{3}{2}}
\label{equ:rss}
\end{eqnarray}

\noindent with $R_{\rm S,0}$ the Str\"{o}mgren radius at $t=t_{\rm
  form}$; $t_{\rm cross} \equiv R_{\rm S,0}/v_{\rm sound}$ the sound
crossing time of the Str\"{o}mgren sphere at $t=t_{\rm form}$; $n_{\rm
  e, 0}$ the electron density at $t=t_{\rm form}$ (assumed to be
$\approx 10^{4}$cm$^{-3}$), and $v_{\rm sound}$ the sound speed in the
gas (assumed to be $\approx$10\kms). 



We then determine the radio flux $S$ at a given
frequency $\nu$ of the \hii-region,

\begin{equation}
S_{\nu} = B_{\nu}(T) [ 1 - \exp(-\tau_{\nu}) ]
\label{equ:radio}
\end{equation}

\noindent where $B_{\nu}(T)$ is the Planck function, we assume a
typical \hii-region temperature of 8000K
\citep[e.g.][]{Dyson-Williams}, and $\tau_{\nu}$ is the optical depth
at frequency $\nu$,

\begin{eqnarray}
\tau_{\nu} =  0.53 ~ \times &
\displaystyle\left(\frac{T}{10^4K}\right)^{-1.35} 
\times \left( \frac{\nu}{GHz}\right)^{-2.1}
\times \nonumber\\
& \displaystyle\left( \frac{n_{\rm e}}{10^{3}cm^{-3}}\right)^{0.66} \times 
\left( \frac{Q_{\rm Lyc}}{10^{49.7}s^{-1}}\right)^{0.33}
\label{equ:tau}
\end{eqnarray}

\begin{figure}
  \centering
  \includegraphics[width=8.cm,bb=100 220 460 740]{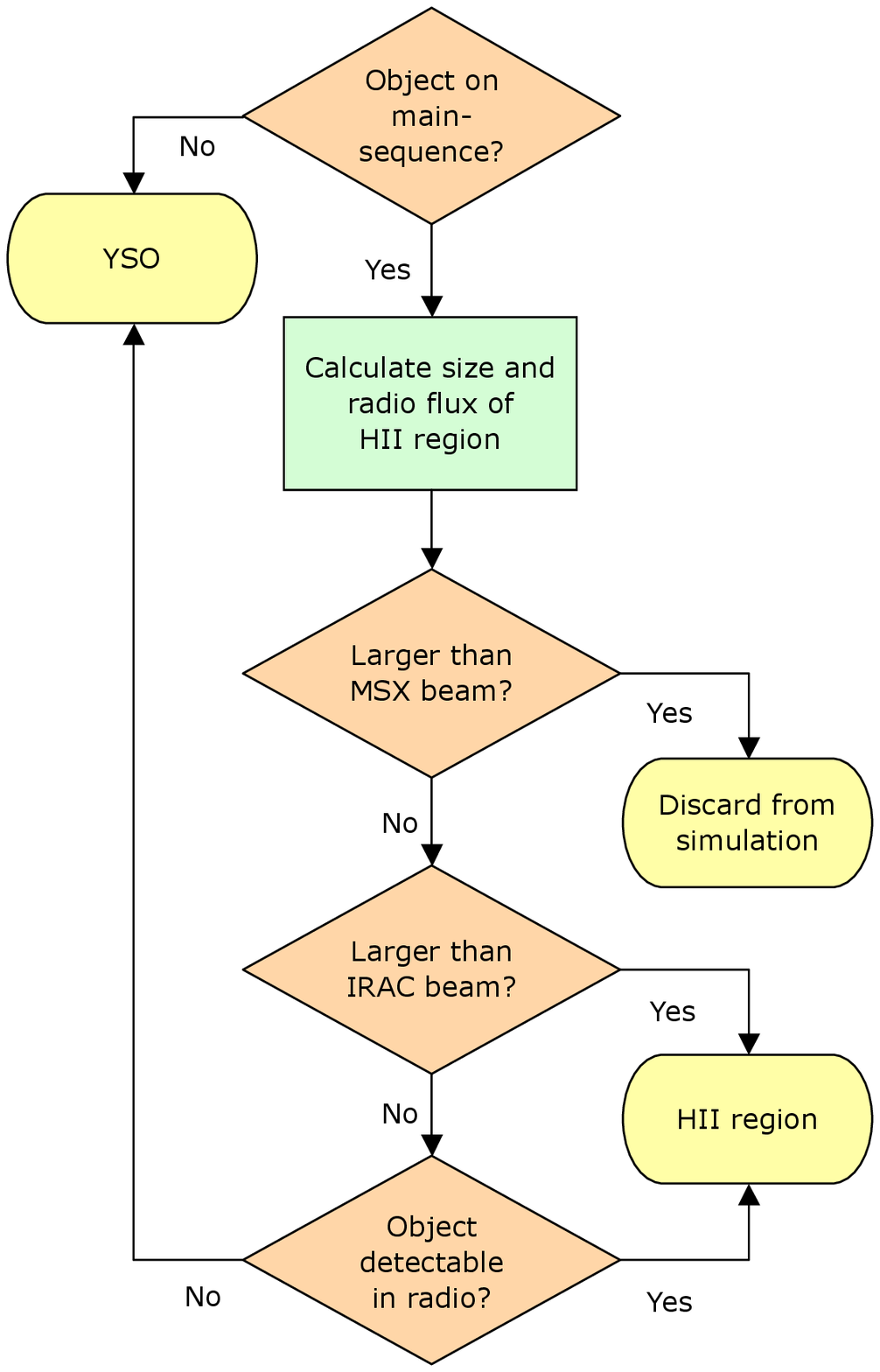}
  \caption{Flow-chart for source classification.}
  \label{fig:flow}
\end{figure}

\subsubsection{Source classification}
Once the observable quantities of each source in the simulation have
been computed, we first apply the detection limits of the \RMS\ survey
to determine whether or not they would form part of the \RMS\ sample. As
the initial source selection was done using MSX point-source fluxes,
the first detection limit to apply is that of the MSX Galactic Plane
survey at 21\microns. As the completeness limits of the MSX survey are
not particularly well defined, we ran our own completeness tests using
sample images from the survey data to calculate the detection limits
as a function of Galactic position. These completeness tests and their
results are described in more detail in Appendix~\ref{sec:msx}. For each
source we compare the 21\micron\ luminosity to the completeness as a
function of 21\micron\ flux at the source's location on the sky, to
determine the probability $P$ that the source would be detected. A
random number $X$ is then generated, and the source is deemed to be
detectable if $X>P$. 

Detected sources are classified as either YSOs or \hii-regions based
on their calculated continuum radio fluxes and angular sizes. Again,
we must then compare these fluxes to the detection limits of the
multiple radio surveys that were used in the compilation of the RMS
survey. If a source is determined to have a radio surface brightness
in excess of the 3$\sigma$-limit along that line-of-sight, it is
classified as a \hii-region. If the size of an object's \hii-region is
greater than the spatial resolution of the imaging part of the
\RMS\ survey (2\arcsec), it is assumed that it would be resolved and
therefore again classified as a \hii-region. If it is greater than the
MSX beam size ($\sim$20\arcsec), it is assumed that the object would
not belong to the MSX point-source catalogue, and so is discarded from
the simulation. Objects which are not detected in the radio, are
point-sources in the high-resolution \RMS\ imaging, and are deemed to
be detectable in the MSX survey, are classified as YSOs. The
classification process is illustrated by the flow-chart in
\fig{fig:flow}.


\section{Results} \label{sec:results}

In the following sections, we will describe the output of the model
with various input parameters. Firstly, we concentrate on the results
when the `fiducial' set of input values are used. 

\begin{table}
  \centering
  \caption{Values of the parameters used in the fiducial model.}
  \label{tab:fid}
  \begin{tabular}{lc}
    \hline \hline
    Parameter & Value \\
    \hline
    Accretion law & \citet{M-T03} \\
    \sigcl        & 1.0 g\,cm$^{-2}$ \\
    $n_{\rm e}$    & 10$^4$cm$^{-3}$ \\
    Gas distribution & Spiral arms only \\
    IMF           & \citet{Kroupa01} \\
    SFR$_{\rm Gal}$ & 3\msunyr \\
    \hline
  \end{tabular}
\end{table}

\subsection{Fiducial model}
To first test the output of the model we start with a simulation for
which the main parameters are set to fiducial values. These values are
listed in Table \ref{tab:fid}. 

For this initial model we start with the accretion law of MT03, with
the parameter \sigcl\ set to 1.0\,g\,cm$^{-2}$. In accord with the
results of MT03, this implies that the mass at which massive stars
join the MS, \mms, is set to 20\msun. The initial electron density
$n_{\rm e}$ when the \hii\ region first turns on is set to
10$^4$cm$^{-3}$, which is a typical value for ultra-compact
\hii-regions \citep{W-C89}. We use a Galactic gas distribution model
in which star formation is confined to the spiral arms {\it only}
(i.e. the `thin-disk' and `thick-disk' diffuse gas components are
ignored). This is because, in the NE2001 model, the thin-disk
component includes a dense circumnuclear ring of gas at a
Galacto-centric distance of \rgc$\approx$5kpc. This produces a large
excess of sources at this location, which is not observed. The spiral
arms on their own however reproduce the spatial distribution of
sources reasonably well (see Sect.\ \ref{sec:gdist}).

\begin{figure*}
  \centering
  \includegraphics[width=8.5cm]{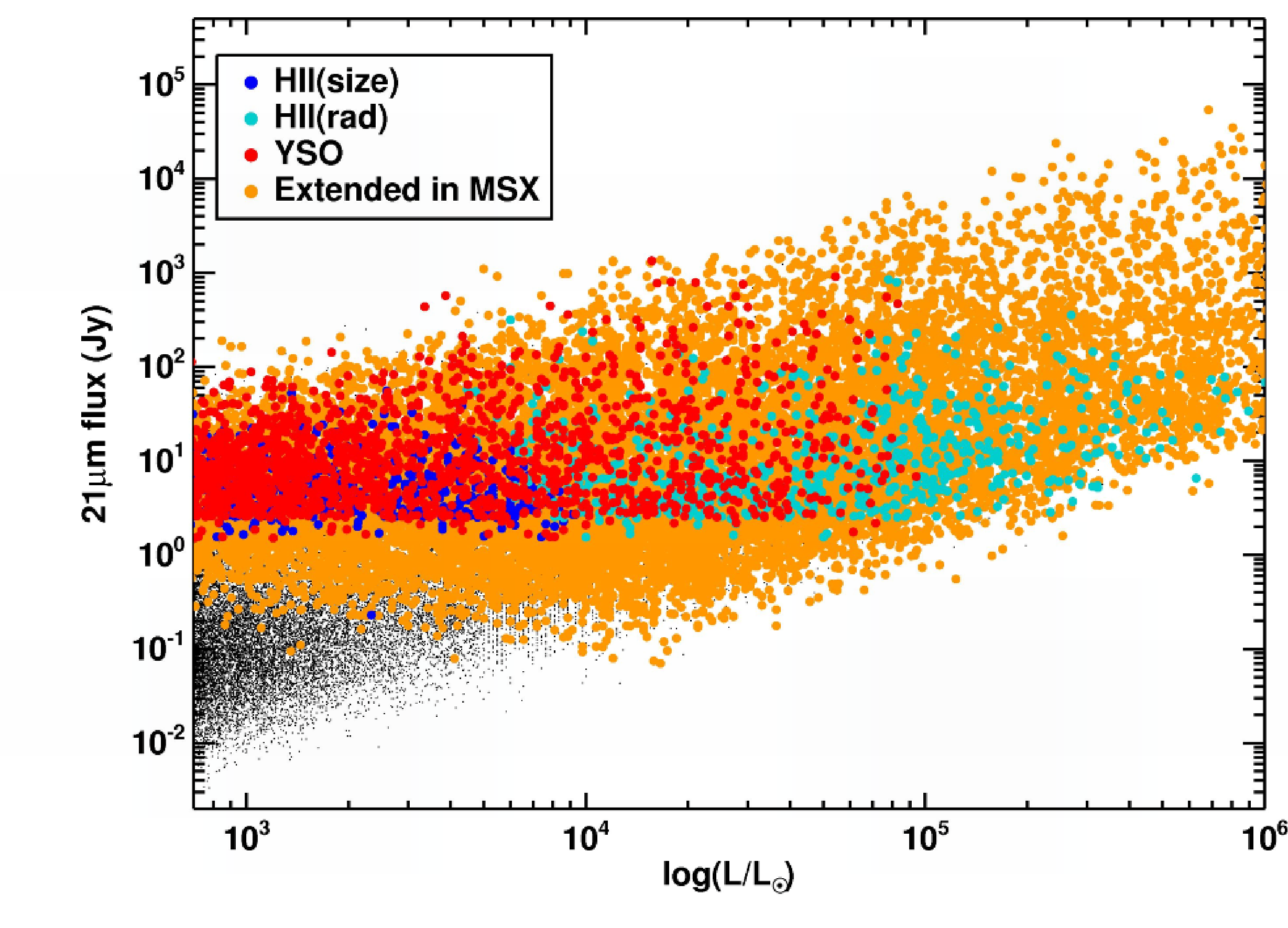}
  \includegraphics[width=8.5cm]{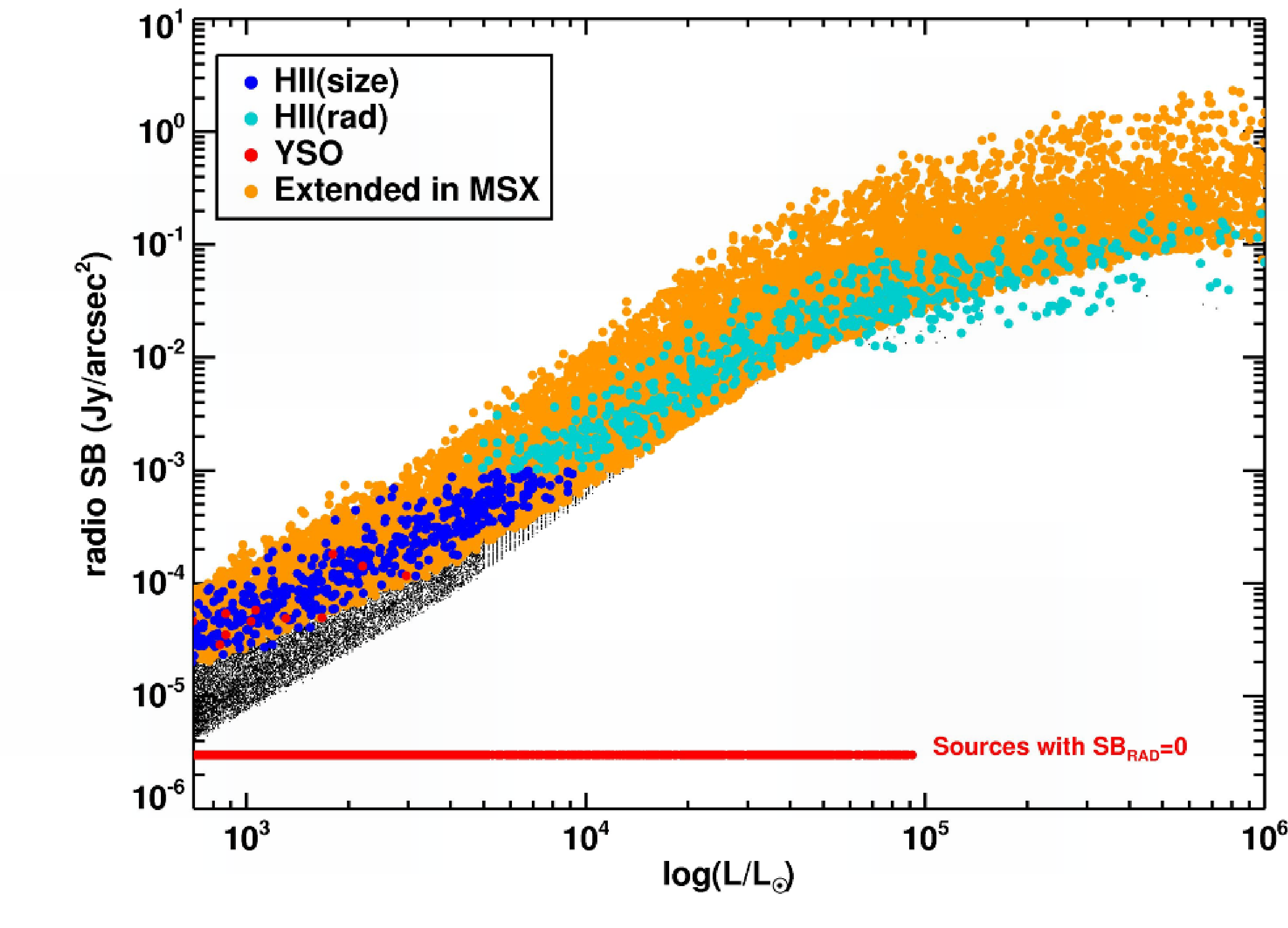}
  \includegraphics[width=8.5cm]{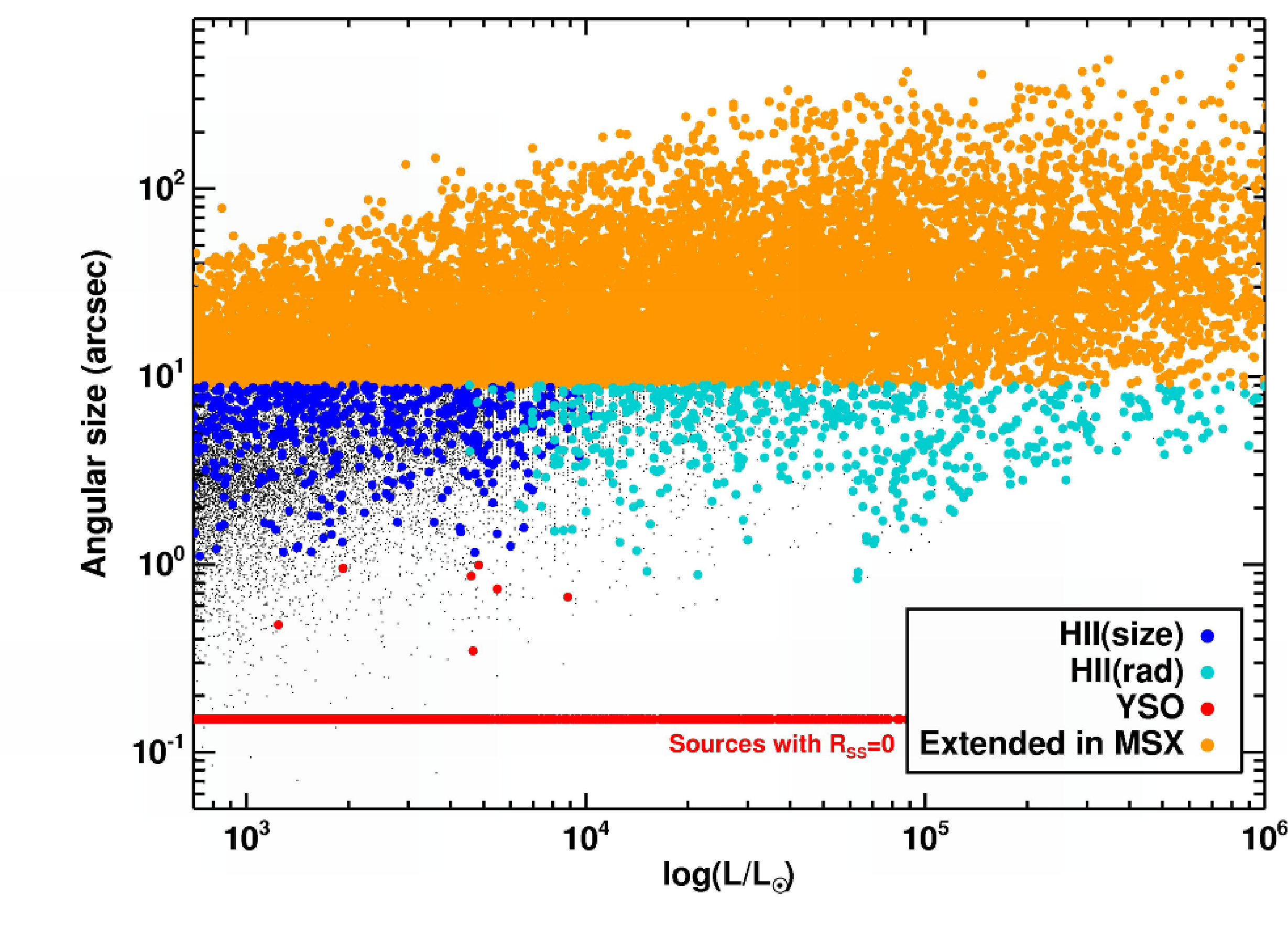}
  \includegraphics[width=8.5cm]{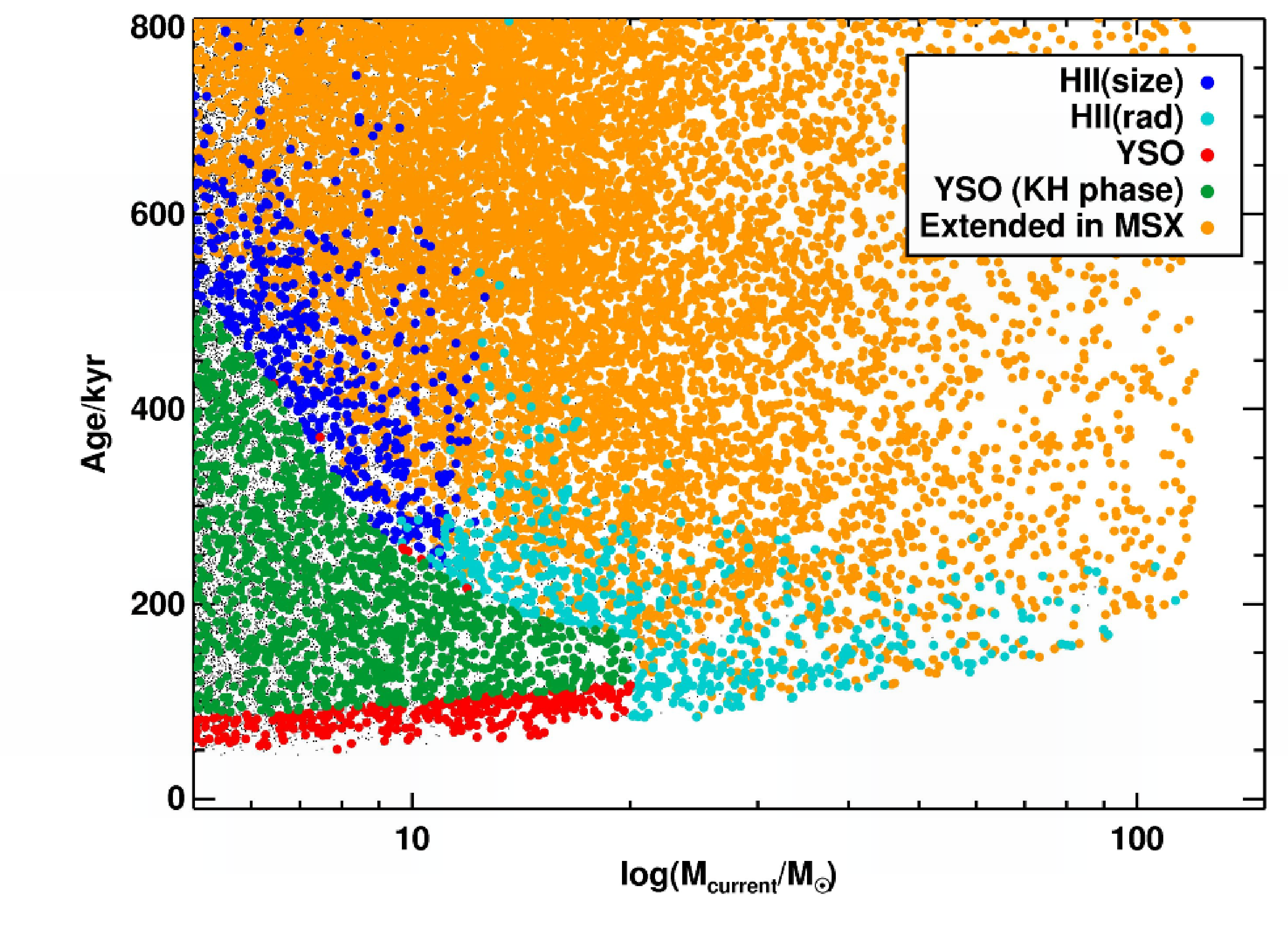}
  \caption{Various physical properties of the stars in a simulation
    using the fiducial parameters. In each panel the symbols have the
    followiing meanings: {\it black-dot}: all objects in the simulation;
    {\it cyan-circles}: objects classified as \hii\ regions based on their
    radio flux; {\it blue-circles}: radio-quiet objects classified as
    \hii\ regions based on their physical sizes; {\it red-circles}: YSOs;
    and {\it yellow-circles}: objects whose apparent sizes are larger than
    the MSX beam and so would be excluded from the RMS survey. The
    four panels show, clockwise from top-left: 21\microns\ flux as a
    function of luminosity; radio surface brightness as a function of
    luminosity; apparent angular size as a function of luminosity; and
    age of each object as a function of its current mass.}
  \label{fig:fiducial}
\end{figure*}

\subsubsection{Illustration of classification criteria} \label{sec:illust}
In \fig{fig:fiducial} we illustrate the effect of our selection
criteria on the output of the model. The top-left panel shows the
21\microns\ flux $F_{21}$ of each object, demonstrating that no
sources below the MSX detection limit ($\sim$4-10\,Jy depending on
location in the plane) are included in catalogue of young stars. The
width and slope of the distribution of $F_{21}$ as a function of
$L_{\rm bol}$ are governed by the empirically derived parameters of
Eq.\,(\ref{equ:f21rat}). Another feature of the simulation illustrated
by this plot is the absence of YSOs above $L_{\rm bol} \approx
10^5$\lsun. This corresponds to the ZAMS luminosity of a star with
\mstar = \mms (=20\msun\ in this simulation). That is, any star which
has reached the MS and is ionizing its nebula is found to be
detectable by the \RMS\ radio observations.

In the top-right panel of \fig{fig:fiducial} we plot the radio surface
brightness of the objects in the simulation. The figure shows that
only sources with surface brightnesses above 1\,mJy/beam are
detectable in the radio, and hence are classified as \hii\ regions
from their radio emission. Below this limit there are further sources
which are classified as \hii\ regions based on their angular size: an
object which is radio-quiet, but that would be found to be extended in
follow-up infrared imaging with spatial resolution
$\sim$2\arcsec\ (i.e. has radius greater than 1\arcsec), is also
classified as a \hii\ region. This is shown again in the bottom-left
panel of \fig{fig:fiducial}, we also illustrate the point at which evolved
\hii\ regions are discarded from the simulation, i.e. when their
angular sizes become larger than 18\arcsec\ and would no longer be
point-sources in MSX. These `discarded' stars are shown as yellow
points.

Another aspect illustrated well by the top-right and bottom-left
panels is that the {\it observational} and {\it physical}
classification criteria match each other extremely well. Massive YSOs
are defined physically as proto-stellar objects which have not yet
contracted to their MS configuration, emit very little Lyman flux, and
so are not surrounded by ionized gas. Meanwhile, the observational
definition is of a proto-star which is bright in the mid-IR but that
shows no evidence of a \hii-region, either from continuum radio
emission of from spatially-extended mid-IR emission. In the top-right
and bottom-left panels of \fig{fig:fiducial} we see that there are very few
objects misclassified as YSOs because their \hii-regions are too small
or weak to be detected. 

\subsubsection{Lifetimes}
The current age of the objects in each phase as a function of current
mass is shown in the bottom right panel of \fig{fig:fiducial}. The objects
classified as YSOs in the simulation all have masses below
20\msun\ ($\equiv$\mms\ in this simulation). That is, as soon as a
massive star reaches the main-sequence and begins to ionize its
surroundings, it becomes immediately detectable in the \RMS\ radio
observations and hence is classified as a \hii-region.

The age distribution of YSOs is a function of current stellar
mass. For objects in the simulation with \mcur$<$20\msun the YSO ages
span 1-3$\times 10^{5}$yrs. Some of these objects are those with high
final masses which are actively accreting. However, most of these YSOs
are objects which have finished accreting are still contracting to the
MS from their `swollen' phase (shown as green points). 

Once arriving on the MS, most objects are then picked up as
\hii-regions. That is, there are very few YSOs outside the `YSO
envelope' clearly defined in the bottom right panel of \fig{fig:fiducial}
by the green and red points (see also discussion in previous
Section). Objects with masses below $\la$10\msun\ have very faint
radio emission, but are still detected as \hii-regions as the angular
size of the gas bubble they drive is larger than the spatial
resolution of the \RMS\ follow-up mid-IR imaging. The \hii-region
phase itself spans around 10$^{5}$yrs for lower mass objects, while it
is about an order of magnitude shorter for the most massive stars.

\begin{figure}
  \centering
  \includegraphics[width=8.5cm]{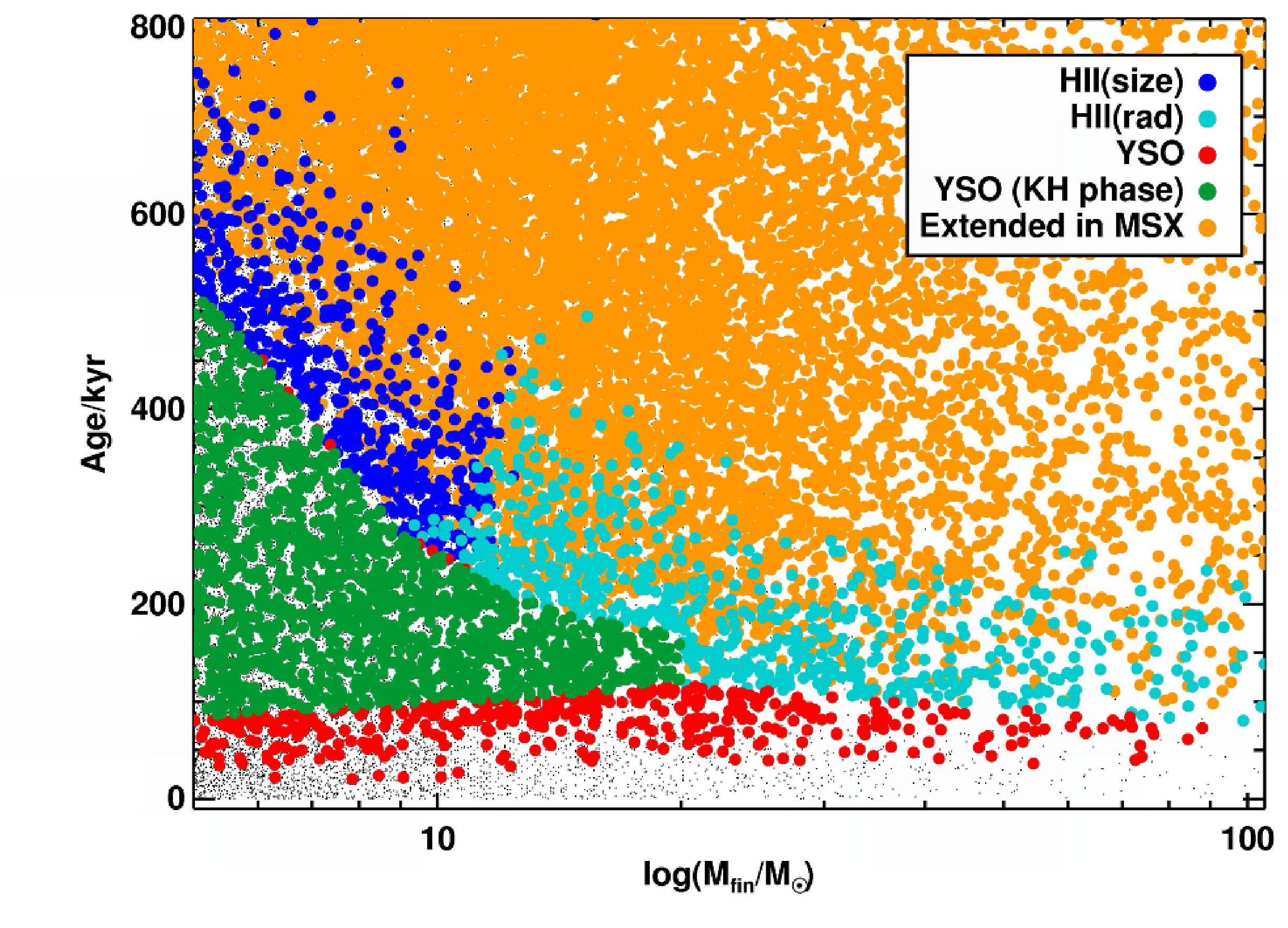}
  \caption{The ages of objects in the simulation as a function of
    their final mass \mfin. Symbols are the same as in the bottom
    right panel of \fig{fig:fiducial}.}
  \label{fig:mfinvage}
\end{figure}

Another illustration of the phase lifetimes is plotted in
\fig{fig:mfinvage}. This plot demonstrates the time spent in each
phase as a function of the {\it final} mass, \mfin. As stars initially
grow in mass very slowly in the MT03 accretion rate model, no objects
in the simulation become visible until around 40,000yrs. The YSO phase
then lasts $\sim 1-3 \times 10^5$yrs for the lowest mass objects, a
substantial fraction of which is spent contracting from the `swollen'
KH phase. For the most massive objects, which can reach the MS while
still accreting, the YSO phase lasts $\sim 5 \times
10^{4}$yrs. Therefore, in these simulations, though the most massive
YSOs are currently $\sim$20\msun, substantial numbers of these objects
are still accreting and will go on to form stars well in excess of
20\msun.

\begin{figure}
  \centering
  \includegraphics[width=8.5cm]{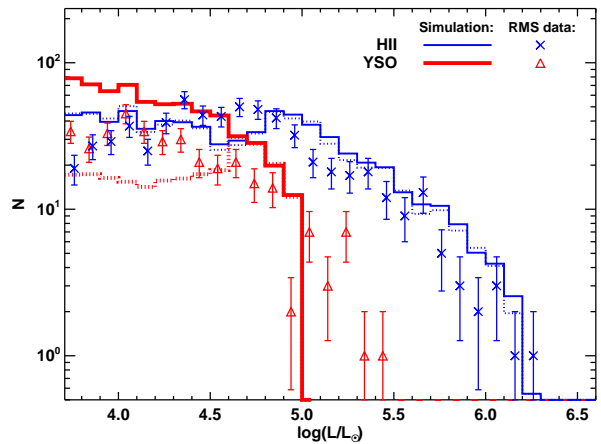}
  \caption{Relative numbers of YSOs and \hii\ regions as a function of
    stellar luminosity. The points with error bars show the results of
    the RMS survey, while the histograms show the results of the
    simulation with fiducial parameters. The thick red solid line and
    the thin blue solid line show YSOs and \hii\ regions respectively
    for the simulation with the fiducial parameters. The dotted lines
    show the results of a simulation in which the Kelvin-Helmholz
    contraction phase has been switched off. The real and simulated
    data have been binned identically, while the luminosities of the
    YSO and \hii\ observations have been offset from one another for
    clarity }
  \label{fig:fid_lfunc}
\end{figure}

\subsection{Comparison to the \RMS\ results}

In the following sections we compare the results of our simulations to
those of the \RMS\ survey. The features that are readily comparable
are the relative distributions of objects as a function of luminosity
and Galactic position. 

Within the \RMS\ survey over 98\% of the objects classified as being
YSOs or \hii-regions now have unambiguous kinematic distances. To
assess the impact of the remaining 2\%, we compared three different
observed luminosity distributions -- one with the ambiguous distances
set to the nearside distance, to the farside distance, and randomized
between near and far. In practice we found that the differences
between these three distributions were within the noise, and so have a
negligible impact on our conclusions. When comparing with our
simulated data, we adopted distances for these objects which were
randomized between near and far.

\subsubsection{Luminosity distribution}
A powerful diagnostic of the \RMS\ database is the relative numbers of
YSOs and \hii-regions as a function of luminosity. The observational
data are shown in \fig{fig:fid_lfunc} as data-points with error bars. It
is clear that the observed distribution of YSOs peaks at a lower
luminosity than does the \hii-region distribution, with many more
\hii\ regions found at higher luminosities \citep[see
  also][]{Mottram11}. 


Overplotted in \fig{fig:fid_lfunc} are the results of our simulation
with fidicial parameters. We first note that our fiducial model
already provides qualitatively good fits to the data. The quantitative
fit to the \hii-region data is also very good, while the simulated YSO
distribution predicts numbers which are typically a factor of $\sim$2
too large. As will be shown later, by fine-tuning the model parameters
we are able to provide excellent fits to the data.

The {\it relative} numbers of YSOs and \hii-regions (i.e.  $N({\rm
  YSO})/N($\hii$)$) are also well matched. This is a validation of the
approximations we made in calculating the observed properties of the
\hii-regions. Under our current assumptions, almost every object in
the simulation which has evolved as far as the MS is detected as a
\hii-region (see also discussion in Sect.\ \ref{sec:illust}). If our
assumptions {\it underestimated} the radio flux $S_{\nu}$ then the
numbers YSOs/\hii-regions would be unchanged, as nearly all MS objects
are already detected. Meanwhile, if we {\it overestimated} $S_{\nu}$
then the \hii-regions would still be picked up as being spatially
extended (see top-right panel of \fig{fig:fiducial}). The physical sizes of
the \hii-regions are goverened by their expansion rates, which in turn
is dictated by the parameter \ne. This can greatly alter the numbers
of \hii-regions in the simulation while leaving the numbers of YSOs
unchanged. Using the typical ultra-compact \hii-region value of
\ne=10$^4$cm$^{-3}$ \citep{W-C89} we find excellent agreement between
the simulated luminosity distributions of YSOs/\hii-regions and the
\RMS\ results.

One obvious feature of the simulated data is the steep drop-off in
YSOs at $\log(L/$\lsun$) = 4.9$, whereas the observations show a 
smoother drop-off at high luminosities. The location of this drop-off
corresponds to the luminosity of objects with large \mfin\ when they
arrive on the MS at \mms. The parameter \mms\ depends on the accretion
rate -- higher accretion rates mean that stars join the MS at higher
masses. Therefore, we can adjust the location of the
high-\lstar\ dropout by changing \mms, or equivalently the accretion
rate.

Finally, we address the topic of the pre-MS `swollen' phase of massive
stars. In \fig{fig:fid_lfunc}, the solid and dotted lines show the
luminosity distributions of simulations with and without a KH
contraction phase respectively. As we have already illustrated in
Figs.\ \ref{fig:fiducial} and \ref{fig:mfinvage} the inclusion of a KH
phase significantly prolongs the YSO phase in lower mass objects, and
results in larger numbers of low-\lstar\ YSOs. In \fig{fig:fid_lfunc},
we see that if we disregard the KH phase the number of
low-\lstar\ YSOs is reduced. Meanwhile, the number of
high-\lstar\ YSOs remains approximately the same, since any KH phase
for these objects would be short compared to the age of the star when
accretion terminates. The numbers of \hii-regions remain unchanged,
since in our simulation the duration of the \hii\ phase is the same
whether or not it is proceeded by a KH phase. These effects result in
a poorer qualitative fit to the data compared to a simulation which
includes a KH phase. We therefore conclude that a KH swollen phase
{\it is} required for stars with masses $\la$20\msun in order for our
simulations to provide a good match to the data.

\begin{figure*}
  \centering
  \includegraphics[width=18cm,bb=35 0 1275 368]{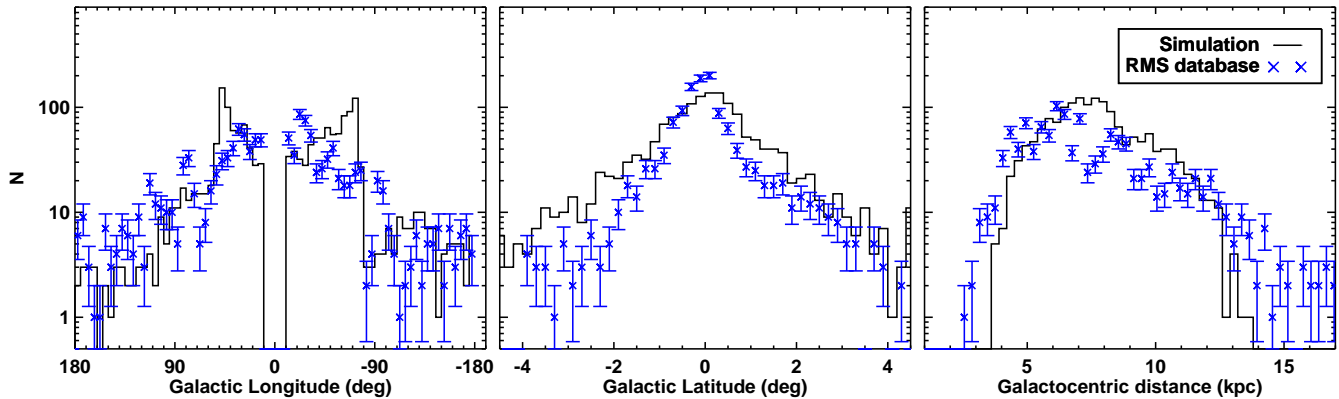}
  \caption{The distribution of YSOs and young \hii\ regions with
    Galactic position. The blue data-points with error bars show the
    results of the RMS survey, while the histograms show the results
    of the simulation using the fiducial model. }
  \label{fig:pos}
\end{figure*}

\subsubsection{Galactic distribution} \label{sec:gdist}
Another diagnostic of the RMS sample is the distribution of sources
with Galactic position. As well as comparing with the source
distribution in Galactic longitude $l$ and latitude $b$, the radial
velocity of each RMS source gives us an unambiguous measurement of
Galacto-centric distance \rgc, assuming that the sources follow the
Galactic rotation curve to within a few \kms. 

We again state that in our fiducial model we consider the star
formation to follow the gas distribution in the {\it spiral arms
  only}. When including the full gas distribution model, which has
components due to the thick disk as well as a ring of gas in the thin
disk at \rgc$\sim$5kpc, we found an extremely large overdensity of
sources at low \rgc\ which is not seen in the data. Our results
suggest that massive star formation is confined to the spiral arm
component of the Galaxy. 

The total distribution of YSOs and young \hii\ regions from the
\RMS\ survey with Galactic position and comparisons with the model
results are shown in \fig{fig:pos}. As with the luminosity
distributions in the previous section, the approximate normalization
and overall trends of the simulations match the data rather
well. However, there are qualitative features of the simulation which
are not matched by the observed data. Firstly, there are two `horns'
in the number counts of the model's $l$-distribution, at $l \approx
50$\degr\ and $l \approx -70$\degr. Secondly, in the $b$-distribution
the fall-off in source counts as one moves away from the plane is
steeper in the observations than in the simulation. Thirdly, the
right-hand panel of \fig{fig:pos} shows that there is a surplus of
sources in the simulation at \rgc$\sim$8kpc with respect to the data.

These three features can all be understood as being due to the same
feature in the Galactic gas distribution model. The horns in the
$l$-distribution are due to an overdensity of sources in the
Sagittarius-Carina arm. As this arm passes close to the Sun, this also
causes the overdensities of sources at high Galactic latitudes and at
\rgc$\sim$8kpc. In principle the \RMS\ results could be used to
fine-tune the description of the Galaxy's gas content, however this is
not the goal of the current work and we leave such a study for a
future paper.

\subsection{Effect of varying fiducial parameters}

\subsubsection{Stellar IMF and {\it SFR}$_{\rm Gal}$}
Broadly speaking, these parameters control the overall normalization
of the predicted number of objects in the \RMS\ survey. The IMF of
\citet{Kroupa01} puts a significant amount of mass into objects with
sub-Solar masses. If this were to be changed to another description of
the IMF, such as a standard Salpeter law down to 0.8\msun, this would
increase the number of massive objects in the simulation, though the
relative numbers of objects as a function of mass would be
unchanged. Though there is speculation as to variations in the slope
of the IMF at high masses and to the presence of an upper-mass cutoff
\citep{Bastian10}, the numbers of objects in the \RMS\ survey with
inferred high masses (i.e. those with \lbol\ $\ga 10^6$\lsun) are too
few for us to test this.

The parameter \SFR\ linearly affects the overall normalization, for
example doubling \SFR\ results in twice as many objects in the
simulation. Changing the IMF to a Salpeter IMF would result in more
massive stars, which would then require \SFR\ to be decreased to
maintain the same overall number counts of objects. However, the
majority of independent estimates of \SFR\ are above 2\msunyr, so
there would be little justification in reducing this parameter beyond
this value. The consistency in overall numbers between our simulations
and the observations serve to justify our initial choices in \SFR\ and
the shape of the IMF. However, later in this study we will allow some
variation in \SFR\ in order to fine-tune the overall normalization.

\subsubsection{Initial \hii\ region density $n_{\rm e}$}
As has already been discussed, changes to this parameter affect only
the duration of the \hii\ region phase. Increasing \ne\ by a factor of
ten from the fiducial value (i.e.\ to \ne=10$^5$cm$^{-3}$) causes the
\hii\ regions to be initially more compact, and expand at a slower
rate. As a result, the number of \hii\ regions in the this model is
increased by a factor of $\sim$5 above those of the fiducial
model. Similarly, dropping the density to \ne=10$^3$cm$^{-3}$ causes
the \hii\ regions to rapidly expand once switched on, shortening the
phase dramatically, leading to an under-prediction of the number of
\hii\ regions again by a factor of $\sim$5. The fiducial value of
\ne=10$^4$cm$^{-3}$ gives a good fit to the observed \hii-region
distribution across all luminosities.

We note that, in reality, \hii-regions are likely to be expanding into
a density {\it gradient}, leading to cometary morpholgies with size
evolutions that differ from those calculated here \citep{A-H06}. Our
use of a constant \ne\ simply represents an average over the early
stages of a \hii-region's evolution.

\subsubsection{Stellar mass -- Q$_{\rm Lyc}$ relation}
As discussed in Sect.\ \ref{sec:lyman}, the relationship between
stellar mass and the ionizing flux \Qlyc\ is uncertain. As has already
been noted by \citet{Martins05} the uncertainty in \Qlyc\ for a star
of a given mass could be in excess of 50\%, though we find from
comparisons of various authors' estimates of the mass -- spectral-type
relation that the uncertainty in \Qlyc\ for a given spectral-type may
be as high as $\sim$0.7dex, or roughly a factor of five.

To investigate the effect of this uncertainty, two further simulations
were run with \Qlyc\ scaled by $\pm$0.7dex. With \Qlyc\ decreased,
\hii\ regions emit less radio flux and expand at a slower rate,
meaning that the \hii\ region phase is prolonged. However, the vast
majority of \hii\ regions are still picked up by the selection
criteria, meaning that very few \hii\ regions are classified as YSOs
in the simulation. The primary effect of decreasing \Qlyc\ is
therefore to increase the number of \hii\ regions, though only by a
factor of $\sim$2 in most luminosity bins.

Similarly, increasing \Qlyc\ by 0.7dex has only a mild impact on the
results. The overall number of \hii\ regions is lowered due to the
shorter \hii\ region lifetime, though again only by a factor of
$\sim$2. So, despite the large uncertainty in the
\Qlyc-\mcur\ relation, we conclude that this uncertainty has
negligible impact on the results of our simulations.


\section{Discussion: a comparison of different accretion laws} \label{sec:accrates}

Thus far, all our models have used the MT03 prescription of accretion,
in which \mcur$\propto M_{\rm fin}^{0.5} t^2$, i.e.\ accretion which
accelerates with time. In the following sections we investigate the
free parameters in this prescription, as well as test other modes of
accretion.

We measure how well a certain accretion law fits the data from the YSO
\lbol-distribution {\it only}. This is because of (a) the large number
of assumptions that have gone in to predicting the \hii-region
properties, and (b) the morphology and normalization of the
\hii-region \lbol-distribution can be augmented by altering parameters
such as \ne\ and \Qlyc, which are uncertain to factors of
$\sim$five. For now, we consider it a success of our model that the
\hii-region \lbol-distribution is matched very well for the fiducial
parameter values.

To assess the results produced by different accretion laws we employ a
standard reduced-\chisq\ analysis between the observed and simulated
\lbol-distributions. We have adaptively rebinned the observed YSO
distribution such that there is a minimum of 9 objects per bin,
i.e. that each bin has a minimum fractional error of 33\% and
therefore has a minimum significance of 3$\sigma$ in Poissonian
statistics. For the simulated data we re-ran each simulation 50 times
to reduce random errors to well below those of the observations. 

We allowed the precise normalization of the \lbol-distribution to
vary, since the parameter that affects overall numbers of objects
(i.e. \SFR) is itself uncertain, and we would not wish to discriminate
against a particular accretion law simply because of a small offset in
numbers. However, where a large normalization factor was required to
achieve a minimal \chisq\ we critically assess the implications for
the associated re-scaling of \SFR. The best-fitting \chisq\ and
associated \SFR\ values for each model run are summarized in Table
\ref{tab:chisq}.

\begin{table}
  \centering
  \caption{Summary of each accretion scenario and associated model
    parameter, its \chisq\ value, and the implied average
    star-formation rate of the Galaxy.}
  \begin{tabular}{lcc}
\hline
Model \& Parameter   &   $\chi^2$   &   \SFR \\
\hline \hline
{\it MT03} \\
\sigcl =  0.1 g\,cm$^2$  &  3.1  &  0.5 \\
\sigcl =  1.0 g\,cm$^2$  &  1.2  &  1.6 \\
\sigcl = 10.0 g\,cm$^2$  &  1.3  &  2.1\smallskip\\
{\it B-H} \\
  \mms = 15     \msun  &  2.3  &  5.5 \\
  \mms = 20     \msun  &  1.0  &  2.0 \\
  \mms = 30     \msun  &  4.0  &  1.8\smallskip\\
{\it Constant $t_{\rm form}$} \\
\tform =  70$\times 10^3$       yrs  &  3.6  &  1.6 \\
\tform = 300$\times 10^3$       yrs  &  2.4  &  0.5 \\
\tform = 700$\times 10^3$       yrs  &  0.7  &  0.6\smallskip\\
{\it Uniform $\dot{M}$} \\
$\log (\dot{M}/M_{\odot}{\rm yr^{-1}})$ = -3   & 41.2 & 7.7 \\
$\log (\dot{M}/M_{\odot}{\rm yr^{-1}})$ = -3.5 & 4.2  & 1.5 \\
$\log (\dot{M}/M_{\odot}{\rm yr^{-1}})$ = -4   & 92.6 & 3.1\smallskip \\
{\it SK04} \\
  \mms = 15     \msun  &  2.0  &  1.5 \\
  \mms = 20     \msun  &  2.1  &  1.2 \\
  \mms = 25     \msun  &  2.8  &  1.1 \\
\hline
  \end{tabular}
  \label{tab:chisq}
\end{table}

\begin{figure}
  \centering
  \includegraphics[width=8.5cm,bb=30 0 708 510]{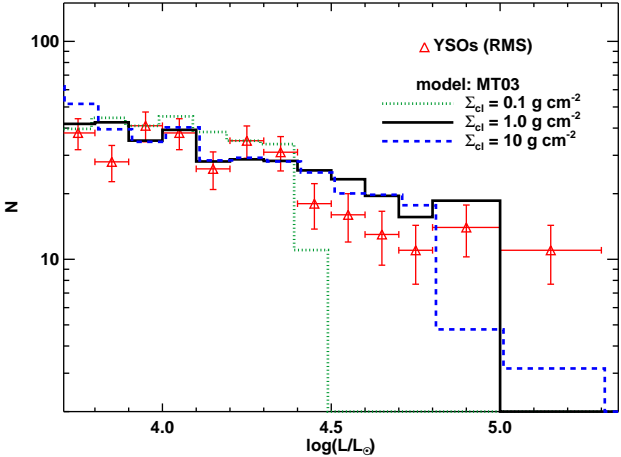}
  \caption{The effect of varying \sigcl, or equivalently the
    time-averaged accretion rate a function of final stellar mass, on
    the simulated luminosity distribution of YSOs. The histograms show
    the simulation results for values of \sigcl=0.1, 1.0 and
    10g\,cm$^{-2}$, which have been renormalized to best match the
    data by scaling the overall star-formation rate (see text for
    details). The data have been adaptively binned to have a minimum
    error per bin of 33\% (i.e. a 3$\sigma$ detection in Poissonian
    statistics).  }
  \label{fig:sigcl}
\end{figure}

\subsection{MT03 accretion}
In the MT03 turbulent core model, the accretion rate of a star of a
given final mass is dictated by the clump pressure and pressure
structure. In their fiducial model this is parameterized in terms of
the clump's surface density \sigcl. Increasing the accretion rate not
only leads to faster formation times, but also results in high
\mfin\ stars joining the MS at higher masses (i.e. \mms\ increases,
due to greater `swelling' in the accretion phase). According to MT03,
varying \sigcl\ between 0.1 and 10g\,cm$^{-2}$ results in values of
\mms\ between 12 and 24\msun. In \fig{fig:sigcl} we show luminosity
distributions using the extrema for these values of \sigcl\ and \mms.

The low-\sigcl\ model has large differences to the fiducial model. The
decreased accretion rates have the effect of lengthening the YSO
phase, which is now dominated by stars still accreting rather than by
stars which have finished accreting and are contracting to the
MS. This means that the total number of low luminosity YSOs in the
simulation is much larger, and to match the observed YSO levels in
\fig{fig:sigcl} we had to normalize the data by reducing the numbers
of objects by a factor of 6. The only physical explanation for this
normalization factor would be to reduce the Galactic star formation
rate to \SFR$\approx$0.5\msunyr, whereas the vast majority of
estimates of this quantity are $>$2\msunyr \citep[see][ and refs
  therein]{Diehl06}. Also, the decrease in \mms\ pushes the YSO
drop-out to lower luminosities, making it much more pronounced, and in
clear disagreement with the \RMS\ results. This is reflected in the
poorer \chisq\ value when compared to other model runs of this
scenario.

The fiducial and high-\sigcl\ models match the observations very
well. The high-\sigcl\ model has a higher value of \mms\ than the
fidicial model, the effect of which is to push the YSO drop-out to
higher luminosities. As the steepness of the IMF dictates that there
are fewer objects at these higher luminosities, this makes the
drop-out slightly less dramatic. The faster accretion rates mean that
massive stars form quicker, however this does not greatly affect the
distributions of YSOs. The YSO population is dominated by those
objects with \mfin$<$\mms, and a substantial fraction of the YSO
lifetime of these objects is the Kelvin-Helmholz contraction phase,
which is independent of the accretion rate in our model. The objects
with \mfin$>$\mms\ spend very little time in the YSO phase in the
fiducial model (cf.\ \fig{fig:mfinvage}), so decreasing the
contribution of such objects to the YSO population does little to the
observed YSO distribution, as the high \mfin\ stars spend most of
their formative years as \hii\ regions.

In terms of which value of \sigcl\ produces the best fit, the fiducial
model achieves the lowest reduced \chisq: the \sigcl=1g\,cm$^{-2}$
model has \chisq=1.2, and the \sigcl=10g\,cm$^{-2}$ has
\chisq=1.4. Given the similarity of these values, and the random
uncertainty associated with how the data are binned, we consider each
of these models to produce equally good fits. However, the fiducial
value of \sigcl\ is comparable to that typically observed in Galactic
star-forming clumps \citep{Plume97}, whereas a value of
\sigcl=10g\,cm$^{-2}$ seems excessively high and is not supported by
observations. The renormalization factor for the \sigcl=1g\,cm$^{-2}$
model is 0.65, implying a global star formation rate of
\SFR=1.6\msunyr. This is reasonably consistent with the vast majority
of estimates of this parameter, which place it in the range
2-4\msunyr. Therefore, we consider that the \sigcl=1g\,cm$^{-2}$ model
provides a good match to the data while being consistent with
complimentary observations of star-forming clumps.

\begin{figure}
  \centering
  \includegraphics[width=8.5cm,bb=30 0 708 510]{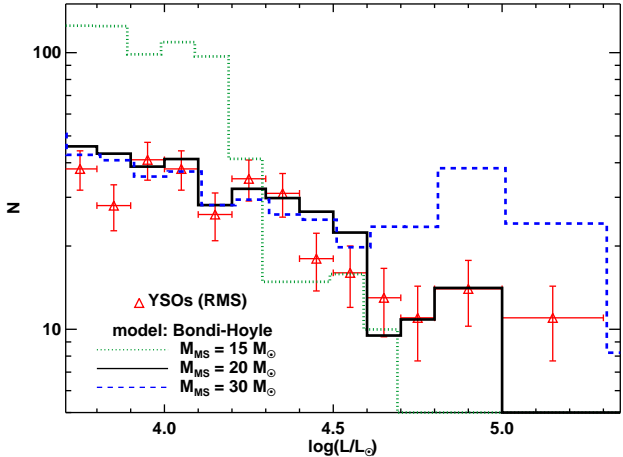}
  \caption{Same as \fig{fig:sigcl} but for Bondi-Hoyle accretion
    models, for three different values of the mass at which accreting
    stars arrive on the main-sequence, \mms. All simulations have been
    re-normalized by a factor of 1.5, see text for details. }
  \label{fig:bh}
\end{figure}

\subsection{Bondi-Hoyle accretion}
The simple Bondi-Hoyle (BH) `competitive' accretion model we are using
has the attractive property that it has very few free parameters once
we employ the normalization of \citet{B-B06}. We are not aware of any
robust numerical simulations of BH accretion which predict properties
such as \mms\ and MS contraction time, so in these simulations we have
assumed that the MS contraction time approximates to the
Kelvin-Helmholz time, in accord with the results of HO09. We allow
\mms\ to remain a free parameter, and explore the range
\mms=15-30\msun\ which are typical values taken from the other
accretion laws studied here. In \fig{fig:bh} we show the results of
these simulations with three different values of \mms\ between 15 and
30\msun. The qualitative agreement with the \RMS\ data is very good
when \mms\ is between 20-30\msun, while the overall normalization is
also very good: optimal renormalization factors were found of
$\sim$0.6, implying \SFR\ between 1.7-2.0\msunyr.

The best fitting model is that with \mms=20\msun, with reduced
\chisq=1.0. This value of \mms\ seems reasonable, given the
quantitative results of MT03 and HO09 who find
\mms=12-30\msun\ depending on the accretion history of the
protostar. The optimal normalization of this model requires
\SFR=1.9\msunyr, which again is reasonable given other measurements of
this value.

\subsection{Constant accretion}

In this section we explore two distinct scenarios. Firstly, we
investigate a uniform accretion rate \mdot\ for all stars, regardless
of mass. Secondly, we fix the accretion rate such that
\mfin/\mdot\ is constant, that is that all stars finish accreting on
the same timescale \tform, independent of \mfin.

\begin{figure}
  \centering
  \includegraphics[width=8.5cm,bb=30 0 708 510]{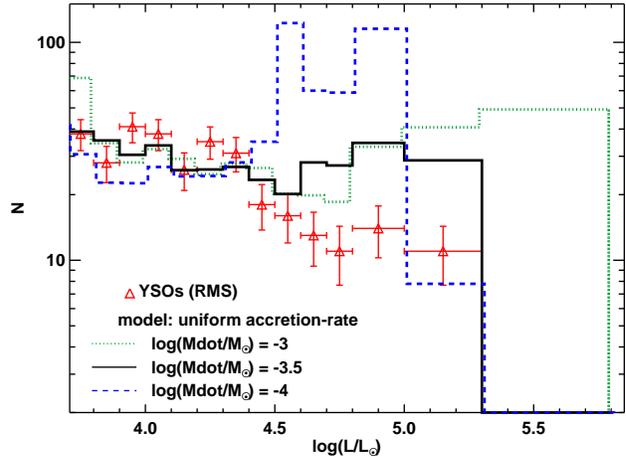}
  \caption{Same as \ref{fig:sigcl} but for uniform constant
    accretion rates. Each histogram has been re-normalized to fit the
    observed numbers of low-\lbol\ YSOs, see text for details.}
  \label{fig:hosu}
\end{figure}

\subsubsection{Uniform \mdot} 
The luminosity distributions for three values of \mdot\ are shown in
\fig{fig:hosu} using the computed tracks at
\mdot=10$^{-3}$\msunyr\ and 10$^{-4}$\msunyr, and a track linearly
interpolated between these two at 10$^{-3.5}$\msunyr. For each track
we found the value of \mms\ by plotting the track on a H-R diagram and
finding the approximate mass at which the track joined the ZAMS of
\citep{Mey-Mae00}. The value of \mms\ was $\approx$27\msun\ for the
$10^{-3.5}$ and $10^{-4}$\msunyr\ tracks, and $\approx$36\msun\ for
the $10^{-3}$\msunyr\ track.

While the normalization is reasonable (see Table \ref{tab:chisq} for
the implied values of \SFR), these models give poor qualitative fits to
the data. The reason for this is that in the HO09 accretion tracks,
the behaviour of \lbol\ with \mcur\ is such that at some point
\lbol\ reaches a temporary plateau. In the simulation, this causes a
pile-up of objects at a certain luminosity. For example, in the
\mdot=10$^{-4}$\msunyr\ this plateau occurs at \lbol$\simeq
10^{4.6}$\lsun, leading to a spike in the luminosity distribution at
this value (see the blue dashed curve in \fig{fig:hosu}). Similarly,
in the \mdot=10$^{-3}$\msunyr\ model, this pile-up occurs at \lbol$\simeq
10^{5.5}$\lsun. The interpolated model (\mdot=10$^{-3.5}$\msunyr) does
not have any such pronounced spikes in the luminosity distribution as
the interpolation process serves to smooth any such spikes
out. However, this model still produces a poor qualitative fit to the
data, the \lbol-distribution being significantly flatter than the
observations.

\begin{figure}
  \centering
  \includegraphics[width=8.5cm,bb=30 0 708 510]{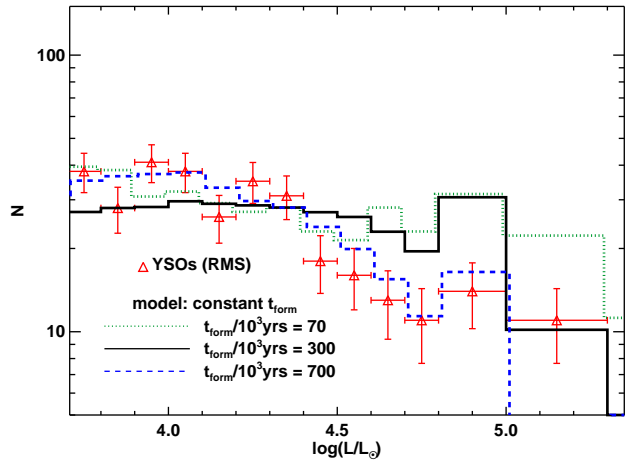}
  \caption{Same as \ref{fig:sigcl} but for uniform formation
    timescales \tform, and with accretion rates which are constant
    with time. The histograms show simulations with \tform\ of $10^5$,
    $3\times 10^5$ and $7 \times 10^5$yrs, and have been multiplied by
    factors of 0.4, 0.1 and 0.2 respectively to fit the observed
    levels. }
  \label{fig:hosc}
\end{figure}

\subsubsection{Uniform \tform}
Here, we experiment with using accretion rates which are constant with
time, but depend on the final mass of the star in such a way that all
stars accrete their matter on identical timescales. To do this, we
took the birthlines of HO09, which are sampled at accretion rates of
10$^{-5}$, 10$^{-4}$ and 10$^{-3}$\msunyr\ and linearly interpolated
them onto a finer grid, sampling at every 0.01dex. For each star in
the simulation we then calculated the accretion rate required for it
to form in a time \tform, where \tform\ is a free parameter, and
assigned each star a constant accretion rate of
\mdot=\mfin/\tform. The lowest values of \mdot\ were clipped at
10$^{-5}$\msunyr, and the highest at 10$^{-3}$\msunyr, as we have
no birthline calculations outside this range. 

The point at which each star joins the MS, \mms, is then a function of
the star's final mass. To determine \mms\ as a function of \mdot\ we
plotted each interpolated track on a H-R diagram and identified the
approximate point at which it joined the ZAMS track of
\citep{Mey-Mae00}. Stars with \mfin$<$\mms\ were again assumed to
undergo a contraction phase before arriving on the MS, with the length
of this phase equal to the K-H timescale of a star of that
mass.

We can summarize the predicted behaviour of such models as
follows. The effect of decreasing \tform\ is to increase the accretion
rates. This means that stars can pass through the YSO phase more
quickly, resulting in less YSOs in each luminosity bin. However, this
effect is mitigated by the fact that the higher accretion rates push
the value of \mms\ to higher masses, extending the length of the YSO
phase for objects with high \mfin. Altering \tform\ can therefore
affect both the qualitative shape and the overall normalization of the
\lbol-distribution.

The results of simulations with three different values of uniform
\tform\ are shown in \fig{fig:hosc}. As expected, in the simulation
with short formation timescales (\tform=$70 \times 10^{3}$yrs) the
YSOs have higher luminosities than in the simulations with longer
\tform. The large value of \mms\ ($\approx$40\msun) means that the
numbers of YSOs with \lbol$\ga 10^{5}$\lsun\ are overpredicted by
factors of 2-3 compared with the lower luminosity objects. At longer
formation timescales of \tform=$3 \times 10^{5}$yrs the YSO drop-out
matches the observations slightly better, though the simulated
distribution is again `top-heavy' compared to the observations. The
simulation with the longest \tform\ produces a very good qualitative
fit to the data. However, the predicted numbers of YSOs are too large
by a factor of five, requiring \SFR\ to be reduced to 0.6\msunyr\ to
match the observed levels.

\begin{figure}
  \centering
  \includegraphics[width=8.5cm,bb=30 0 708 510]{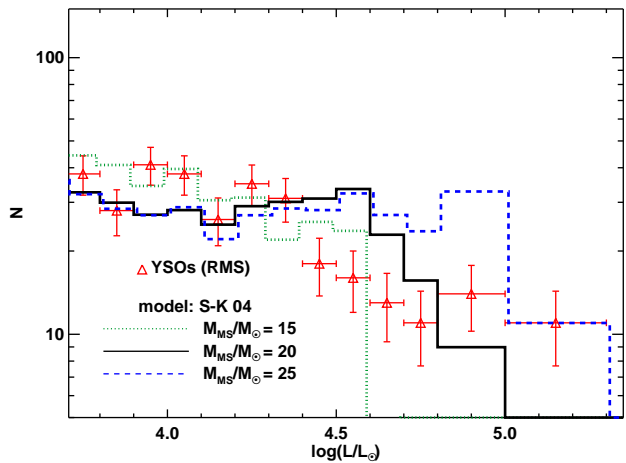}
  \caption{Same as \ref{fig:sigcl} but with accretion rates from
    SK04 which decelerate with time. Simulations are shown using three
    different values of \mms. The simulations have been renormalized
    to fit the data by reducing the total number of objects in the
    simulation by a factor of three. }
  \label{fig:sk04}
\end{figure}

\subsection{Decelerating accretion}

As already discussed in Sect.\ \ref{sec:accrates}, the prescription of
decelerating accretion from \citet{S-K04} cannot produce massive stars
within a reasonable formation timescale when using the standard values
for the constants in Eq.\ (\ref{equ:sk04}). Massive stars can be
formed however if the constants are pushed to the boundaries of their
uncertainties. This maximises the value of $\tau$ in
\eq{equ:sk04_tau}, and allows the protostars to spend more time at
high accretion rates. For the analysis in this Section, we use values
of $\tau_{0} = 4.9 \times 10^{4}$yrs and $\tau_{1} = 6.3 \times
10^{4}$yrs.

\Fig{fig:sk04} shows the luminosity distribution of young massive
stars from the simulation using the decelerating accretion
model. The SK04 accretion model provides no estimate of \mms, but the
low accretion rates as the star is approaching its final mass suggest
that \mms\ may be lower under this model than for the constant and
accelerating models. In \fig{fig:sk04} we show
simulation results for three different values of \mms. 

The qualitative shape of the YSO \lbol-distribution this time does not
match the observations. The longer time that YSOs spend at higher
masses compared to the accelerating accretion models produces a
luminosity distribution which is more top-heavy than the
observations. This is reflected in the \chisq\ values, which are
higher than those of the accelerating accretion models. 

The overall numbers of objects, as with the constant rate accretion
law, are overpredicted in this model -- to achieve the correct
normalization each histogram had to be scaled down by factors of 2-3,
implying global star-formation rates below 1-1.5\msunyr. Meanwhile,
the qualitative shape of the models' \lbol-distributions is much
flatter than the observations. This is due to the very high accretion
rates that objects have at early times, and hence higher accretion
luminosities. The net effect is that lower mass objects are moved to
higher luminosity bins with respect to accelerating accretion models,
producing a flatter \lbol-distribution.

\subsection{Which accretion laws fit the data best?}
We now discuss which of the accretion models we have tested best fits
the observed luminosity distribution of the \RMS\ survey data, taking
into account the results described in this section thus far. 

The first conclusion that we draw is that the decelerating accretion
scenario is the least likely of all those studied here. While it may
be argued that the prescription of SK04 is only calibrated between
1-10\msun, the poor match between the data and the observations is
still evident at low luminosities (and hence at masses between 5 and
20\msun). Also, irrespective of the prescription's quantitative
details, the qualititave effect of a decelerating accretion law is to
produce a flat luminosity distribution of YSOs. Therefore, we conclude
that this mechanism is unlikely to be the prevalent mode of massive
star formation in our Galaxy. There are however caveats to this
statement as we made several approximations in our calculations, such
as that of the accretion luminosity of massive protostars, and of
objects joining the MS mid-way through their accretion phase just as
the accelerating accretion models do. Stronger critical analysis of
the decelerating accretion scenario awaits the numerical predictions
of the accretion luminosities and pre-MS tracks of massive stars
forming in this way.

Similarly, we also argue against the uniform accretion rate
models. Models where all stars accrete at exactly the same rate
produce spikes in the luminosity distribution due to the discrete
features in the HO09 accretion tracks, while the global gradients of
the simulated \lbol-distributions are much flatter than that observed.

\begin{figure}
  \centering
  \includegraphics[width=8.5cm,bb=30 0 708 510]{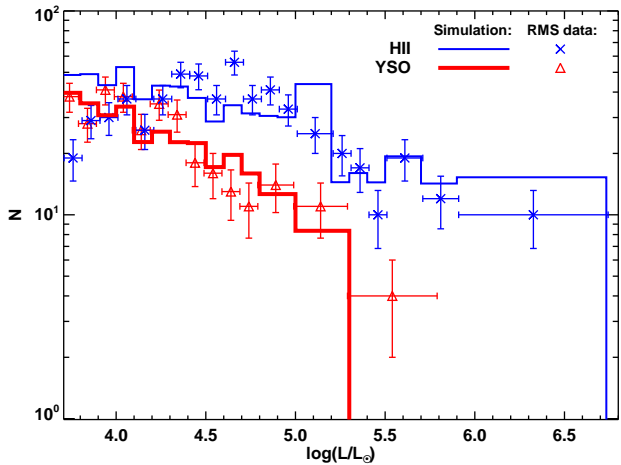}
  \caption{Same as Fig.\ \ref{fig:fid_lfunc} but with the best-fit model, using
    the BH accretion law with \mms=23\msun. To match the \hii-regions,
  we used an initial electron density of \ne=$4\times 10^4 {\rm
    cm^3}$. }
  \label{fig:bestfit}
\end{figure}

The uniform \tform\ models do produce very good qualitative fits to
the data, but only if the formation timescale is very long, i.e. $\sim
7 \times 10^5$yrs. As previously discussed, using the fiducial value
of \SFR\ this produces greater numbers of YSOs than are observed. To
recover the observed numbers, it requires that \SFR\ be reduced to
0.6\msunyr, a factor of five lower than the typically-quoted value of
3\msunyr, and below even the lower limit of \citet{R-W10}, whose
estimate of \SFR\ is significantly lower than any other recent
measurements. Further, this formation timescale would imply
accretion-rates for massive stars of \mdot$\la 10^{-4}$\msunyr, which
are an order of magnitude lower than current observational and
theoretical estimates \citep[e.g.][]{Krumholz09,Qiu11}.

The best fits are produced by the two accelerating accretion
models. Both produce reduced \chisq\ statistics close to unity (see
Table \ref{tab:chisq}), and hence provide excellent fits to the data
given the experimental uncertainties. At the same time, the number of
sources in each model is a close match to the observations, meaning
that the implied \SFR\ remains sensible: we find best fitting average
Galactic star-formation rates of 1.6\msunyr\ and 2.0\msunyr\ for the
MT03 and BH models respectively. These are consistent with most recent
measurements of this property \citep[see][ and references
  therein]{Diehl06,R-W10}.

At this stage, we are unable to conclusively state which of the
accelerating models fits the data best. Random errors in, for example,
how the data are binned, mean that the difference in the \chisq
statistic between the two models is not significant. We should state
that the MT03 model has a more rigorous quantitative basis compared to
our BH model, which simply assumes the analytical formula for
Bondi-Hoyle accretion and ad-hoc estimates for \mms. Thorough
numerical calculations of each scenario may in future allow us to
distinguish decisively between the two.

In \fig{fig:bestfit} we plot the YSO and \hii-region luminosity
distributions for the best-fit model. We have chosen the BH model,
since this gives a slightly lower \chisq\ than the MT03 model, and we
have increased \mms\ to 23\msun\ to match the largest YSO luminosity
bin, though this is more for aesthetic purposes since the reduced
\chisq\ is no better than for \mms=20\msun. Since the overall level of
\hii-regions was slightly underpredicted, we have increased the
initial electron density to \ne=$4\times 10^4 {\rm cm^3}$ to slightly
prolong the \hii-region phase and therefore increase the overall
number of objects. Note that this value of \ne\ is consistent with the
typically quoted densities for ultra-compact \hii-regions of \ne$\ga
10^4{\rm cm^3}$ \citep[][]{W-C89}.

We can highlight two features of \fig{fig:bestfit} where there is room
for improvement in the fit. Firstly, the simulated YSO cutoff occurs
at lower luminosities than in the observed trend. A better fit may be
obtained by using values of \mms\ which depend on \mfin\ rather than a
single blanket value. Since \mms\ depends on accretion history (see
HO09, plus discussion in this work), which in turn depends on stellar
mass in the accelerating accretion rate models, it is entirely
plausible that \mms\ may be dependent on final stellar mass. This
would then serve to smooth-out the YSO cutoff and make it more like
the observations.

Secondly, there is an apparent downturn in the \hii-region luminosity
distribution at low luminosities, which is not reproduced in our
simulations. An improvement to the fit may be obtained if lower
\mfin\ objects have lower initial gas densities once they have reached
the MS. Since low density regions expand at a faster rate, this will
cause the \hii-region phase lifetime to be shorter for these
objects. Though there are a large number of assumptions that go into
our simulation of the \hii-region distribution, it is not an
unreasonable expectation that higher mass stars should form in regions
of higher gas densities.

\begin{figure}
  \centering
  \includegraphics[width=8.5cm,bb=30 0 708 510]{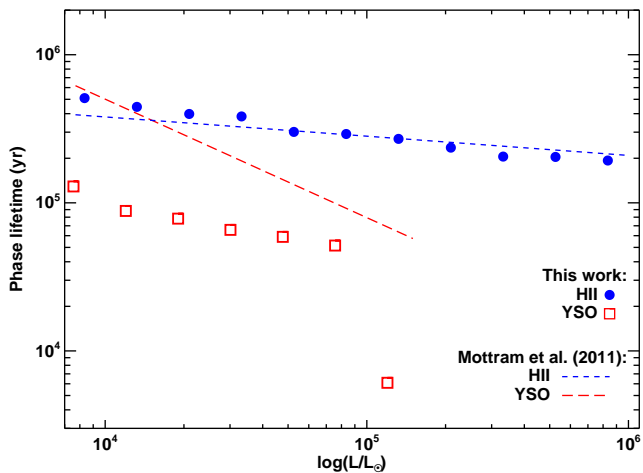}
  \caption{The phase lifetimes, predicted by the best-fit model, for
    both YSOs and \hii-regions, as a function of \lbol. The dotted and
    dashed lines indicate the lifetimes derived empirically by
    \citet{Mottram11L}. }
  \label{fig:lifetimes}
\end{figure}

\subsection{Phase lifetimes of the best-fit model}
For our best-fitting model described in the previous Section, we now
revisit the topic of the YSO and \hii-region phase lifetimes. For
this, we compare to our recent empirical estimates presented in
\citet{Mottram11L}. In that study, the lifetimes were estimated by
dividing the number of objects per luminosity bin by the number of OB
stars with the same luminosity, then multiplying by the MS lifetime of
OB stars. There are three main sources of uncertainty in estimating
phase lifetimes in this way. Firstly, the OB star population of the
Galaxy is only complete within a few kiloparsecs of the Sun, which
serves to increase the random errors. Secondly, the relationship
between spectral type, stellar luminosity and stellar mass is still
uncertain, and may introduce systematic errors, though we mitigate
this uncertainty by using the same sources for these
relations. Finally, this method gives phase lifetime as a function of
{\it luminosity}, which for actively accreting objects is variable
throughout their formation. However, our simulation provides a
`snap-shot' of the Galaxy's population of young massive stars in a way
designed to mimic the \RMS\ survey. Therefore, we {\it can} make a
quantitative measurement of lifetime versus \lbol\ from our
simulation, despite the fact that objects may migrate from bin to bin
throughout their evolution.

To calculate the lifetime as a function of luminosity, we first
construct a diagnostic diagram of the simulation similar to that of
the lower-left panel in \fig{fig:fiducial}, but plotting age against
\lbol. Then, for a given luminosity bin, we determine the age {\it
  range} (i.e. the difference between the minimum to maximum ages) of
all YSOs/\hii-regions in that bin, discarding the oldest and youngest
5\% of objects to limit the impact of statistical outliers. The
results of this analysis are plotted in \fig{fig:lifetimes}, where we
also compare with the results of \citet{Mottram11L} which are
illustrated by the long- and short-dashed lines. 

We find that the lifetimes of YSOs are typically $\sim
10^5$yrs with shorter lifetimes for the more luminous objects,
predominantly an effect of the reduction in KH timescales at higher
masses. On the other hand, \hii-region lifetimes show a weaker
dependence on luminosity, spanning a factor of $\sim$3 between across
the plotted luminosity range\footnote{Note that this lifetime refers
  to that of compact phase of the \hii-region, such that it would be
  detected in the \RMS\ survey. }. This can easily be understood from
a numerical analysis of Eqs.\ (\ref{equ:rss0}) and (\ref{equ:rss}) and
the Lyman fluxes from Table \ref{tab:logq} (see
Sect.\ \ref{sec:radio}), which show that the expansion rates of
\hii-regions span a similar factor of $\sim$3 across the range of
luminosities plotted in \fig{fig:lifetimes}, under the assumption of
initial gas densities which are uniform across all objects.

The agreement between the two estimates of \hii-region lifetimes is
excellent. Recall that we have already fine-tuned the initial electron
density \ne\ in order to match the observed {\it numbers} of
\hii-regions. Since the observed number of objects at any given time
is a product of the production rate and the lifetime, the fact that we
can match the lifetimes and overall numbers simultaneously must mean
that we also have an accurate estimate of the production rate
(i.e.\ the globally averaged star-formation rate). In other words,
the overall number of \hii-regions depends on \SFR\ and \ne, whereas
the lifetime depends on only \ne. 

In terms of the YSOs, there is an apparent discrepancy between the two
estimates. The systematic errors quoted by \citet{Mottram11L},
$\pm$0.45dex in the zero-point and $\pm$0.5 in the gradient, mean that
the differences are significant only at the 1-2$\sigma$ level. In
addition, the differences in gradient can in part be explained by the
way in which \citet{Mottram11L} determined the YSO lifetimes. As
discussed above, these authors assumed a 1:1 correspondance between
the luminosities of the YSOs and the those of main-sequence (MS)
stars. However, since YSOs are often accreting objects they will have
some accretion luminosity, and so their total luminosities will exceed
those of MS stars with the same mass. In contrast, the majority of
\hii-regions in our simulation have finished accreting, and have
bolometric luminosities comparable to their MS luminosities.

For the simulation used in \fig{fig:lifetimes}, we found that if the
YSOs bolometric luminosities were used to estimate \mcur\ (i.e from
interpolation of Table \ref{tab:logq}), then the current masses were
typically overestimated by around 30\%. This leads to the ratio of the
number of YSOs to the number of MS stars at a given mass being
over-predicted, and ultimately to overestimates of the YSO
lifetimes. Quantitatively, we find that this overestimate in lifetimes
is $\sim$70\% at \lbol=$10^4$\lsun, and 40\% at
\lbol=$10^5$\lsun. Though this effect is not large enough to fully
explain the discrepancy in YSO lifetimes in \fig{fig:lifetimes}, it
may be a contributing factor to the difference between the two
gradients.


\section{Summary} \label{sec:conc}
We have constructed a model to simulate the distributions and physical
properties of massive protostars throughout the Galaxy, with the aim
of predicting the observed luminosity distribution of massive Young
Stellar Objects (YSOs) from the \RMS\ survey. To compute the observed
properties of each protostar in the simulation we employed several
different prescriptions of the rates at which stars accrete mass as a
function of time. For each accretion model we identified the
parameters which produced the best qualitative fit to the observed YSO
\lbol-distribution, allowing the global star formation rate \SFR\ to
vary to optimize the overall normalization.

Our main findings may be summarized as follows:

\begin{itemize}
\item The luminosity distribution of YSOs is best described using
  accretion rates which increase as the star grows in
  mass. Accretion rates which were constant in time or which
  decreased as the star grew in mass were ruled out, as they either
  produced poor qualitative fits, produced far too many YSOs, or both.
\item The lack of YSOs at high luminosities is consistent with a
  picture of star-formation whereby stars arrive on the main-sequence
  (MS) once they reach a mass of 20-30\msun, even if the final mass of
  the star is well in excess of this value. We suggest that the
  precise mass at which stars arrive on the MS may be some function of
  the final stellar mass.
\item In order to produce a satisfactory qualitative fit to the YSO
  distribution at low luminosities, an intermediate radio-quiet phase
  is required between the end of accretion and the ignition of a
  \hii-region. The length of this phase is well described by the
  star's Kelvin-Helmholz timescale, and therefore is consistent with a
  `swollen star' period of pre-MS evolution for stars with masses
  $\sim$5-20\msun.
\item The maximum final stellar mass \mfin\ at which stars no longer
  experience this `swollen' phase is indicated by the observed
  drop-off in YSO numbers at $L_{\rm bol} \sim 10^{5.5}$\lsun. This
  corresponds to stars with masses \mfin$\ga$20\msun, which in our
  simulations have accretion rates of 1-4 $\times
  10^{-4}$\msunyr.
\item From the best-fit models we find that the overall numbers of
  objects can be well reproduced by a globally averaged star-formation
  rate of the Galaxy of 1.5-2\msunyr. 
\item Our best-fit model predicts phase lifetimes for YSOs of $\sim
  10^5$yrs, falling off dramatically for objects with \lbol$\ga
  10^5$\lsun. The compact \hii-regions in the \RMS\ survey have phase
  lifetimes between 2-4$\times 10^5$yrs.
\end{itemize}

In the future, the framework of the simulations presented here could
be adapted to investigate Galactic structure, once the distances to
the \RMS\ survey objects are known to a greater precision.  Future
modelling of the RMS population will take into account the evolution
of the SED. This will include the increasing luminosity and heating as
well as the dispersal of circumstellar material by feedback.  Bipolar
cavities in particular make the SED dependent on the viewing angle
\citep{Whitney03} and a simulation of a large sample with random
viewing angles will be constrained by comparison with the colour
distribution as well as properties of the CO outflows and near-IR
reflection nebulae.


\section*{Acknowledgments}
We thank Jonathan Tan and Willem-Jan de Wit for many useful
discussions during the course of this work. BD is funded by a
fellowship from the Royal Astronomical Society.

\bibliographystyle{/fat/Data/bibtex/apj}
\bibliography{/fat/Data/bibtex/biblio}

\begin{thebibliography}{71}
\expandafter\ifx\csname natexlab\endcsname\relax\def\natexlab#1{#1}\fi

\bibitem[{{Arthur} \& {Hoare}(2006)}]{A-H06}
{Arthur}, S.~J. \& {Hoare}, M.~G. 2006, \apjs, 165, 283

\bibitem[{{Ascenso} {et~al.}(2007){Ascenso}, {Alves}, {Vicente}, \&
  {Lago}}]{Ascenso07}
{Ascenso}, J., {Alves}, J., {Vicente}, S., \& {Lago}, M.~T.~V.~T. 2007, \aap,
  476, 199

\bibitem[{{Bastian} {et~al.}(2010){Bastian}, {Covey}, \& {Meyer}}]{Bastian10}
{Bastian}, N., {Covey}, K.~R., \& {Meyer}, M.~R. 2010, \araa, 48, 339

\bibitem[{{Benjamin} {et~al.}(2005){Benjamin}, {Churchwell}, {Babler},
  {Indebetouw}, {Meade}, {Whitney}, {Watson}, {Wolfire}, {Wolff}, {Ignace},
  {Bania}, {Bracker}, {Clemens}, {Chomiuk}, {Cohen}, {Dickey}, {Jackson},
  {Kobulnicky}, {Mercer}, {Mathis}, {Stolovy}, \& {Uzpen}}]{Benjamin05}
{Benjamin}, R.~A., {Churchwell}, E., {Babler}, B.~L., {Indebetouw}, R.,
  {Meade}, M.~R., {Whitney}, B.~A., {Watson}, C., {Wolfire}, M.~G., {Wolff},
  M.~J., {Ignace}, R., {Bania}, T.~M., {Bracker}, S., {Clemens}, D.~P.,
  {Chomiuk}, L., {Cohen}, M., {Dickey}, J.~M., {Jackson}, J.~M., {Kobulnicky},
  H.~A., {Mercer}, E.~P., {Mathis}, J.~S., {Stolovy}, S.~R., \& {Uzpen}, B.
  2005, \apjl, 630, L149

\bibitem[{{Berdnikov} {et~al.}(2000){Berdnikov}, {Dambis}, \&
  {Vozyakova}}]{Berdnikov00}
{Berdnikov}, L.~N., {Dambis}, A.~K., \& {Vozyakova}, O.~V. 2000, \aaps, 143,
  211

\bibitem[{{Bernasconi} \& {Maeder}(1996)}]{B-M96}
{Bernasconi}, P.~A. \& {Maeder}, A. 1996, \aap, 307, 829

\bibitem[{{Bibby} {et~al.}(2008){Bibby}, {Crowther}, {Furness}, \&
  {Clark}}]{Bibby08}
{Bibby}, J.~L., {Crowther}, P.~A., {Furness}, J.~P., \& {Clark}, J.~S. 2008,
  \mnras, 386, L23

\bibitem[{{Blanc} {et~al.}(2009){Blanc}, {Heiderman}, {Gebhardt}, {Evans}, \&
  {Adams}}]{Blanc09}
{Blanc}, G.~A., {Heiderman}, A., {Gebhardt}, K., {Evans}, N.~J., \& {Adams}, J.
  2009, \apj, 704, 842

\bibitem[{{Bonnell} \& {Bate}(2006)}]{B-B06}
{Bonnell}, I.~A. \& {Bate}, M.~R. 2006, \mnras, 370, 488

\bibitem[{{Brandner} {et~al.}(2008){Brandner}, {Clark}, {Stolte}, {Waters},
  {Negueruela}, \& {Goodwin}}]{Brandner08}
{Brandner}, W., {Clark}, J.~S., {Stolte}, A., {Waters}, R., {Negueruela}, I.,
  \& {Goodwin}, S.~P. 2008, \aap, 478, 137

\bibitem[{{Chapman} {et~al.}(2009){Chapman}, {Mundy}, {Lai}, \&
  {Evans}}]{Chapman09}
{Chapman}, N.~L., {Mundy}, L.~G., {Lai}, S.-P., \& {Evans}, N.~J. 2009, \apj,
  690, 496

\bibitem[{{Cordes} \& {Lazio}(2002)}]{C-L02}
{Cordes}, J.~M. \& {Lazio}, T.~J.~W. 2002, ArXiv Astrophysics e-prints

\bibitem[{{Davies} {et~al.}(2009){Davies}, {Figer}, {Kudritzki}, {Trombley},
  {Kouveliotou}, \& {Wachter}}]{SGR1900paper}
{Davies}, B., {Figer}, D.~F., {Kudritzki}, R., {Trombley}, C., {Kouveliotou},
  C., \& {Wachter}, S. 2009, \apj, 707, 844

\bibitem[{{Davies} {et~al.}(2007){Davies}, {Figer}, {Kudritzki}, {MacKenty},
  {Najarro}, \& {Herrero}}]{RSGC2paper}
{Davies}, B., {Figer}, D.~F., {Kudritzki}, R.-P., {MacKenty}, J., {Najarro},
  F., \& {Herrero}, A. 2007, \apj, 671, 781

\bibitem[{{Davies} {et~al.}(2008){Davies}, {Figer}, {Law}, {Kudritzki},
  {Najarro}, {Herrero}, \& {MacKenty}}]{RSGC1paper}
{Davies}, B., {Figer}, D.~F., {Law}, C.~J., {Kudritzki}, R.-P., {Najarro}, F.,
  {Herrero}, A., \& {MacKenty}, J.~W. 2008, \apj, 676, 1016

\bibitem[{{Diehl} {et~al.}(2006){Diehl}, {Halloin}, {Kretschmer}, {Lichti},
  {Sch{\"o}nfelder}, {Strong}, {von Kienlin}, {Wang}, {Jean}, {Kn{\"o}dlseder},
  {Roques}, {Weidenspointner}, {Schanne}, {Hartmann}, {Winkler}, \&
  {Wunderer}}]{Diehl06}
{Diehl}, R., {Halloin}, H., {Kretschmer}, K., {Lichti}, G.~G.,
  {Sch{\"o}nfelder}, V., {Strong}, A.~W., {von Kienlin}, A., {Wang}, W.,
  {Jean}, P., {Kn{\"o}dlseder}, J., {Roques}, J., {Weidenspointner}, G.,
  {Schanne}, S., {Hartmann}, D.~H., {Winkler}, C., \& {Wunderer}, C. 2006,
  \nat, 439, 45

\bibitem[{{Diolaiti} {et~al.}(2000){Diolaiti}, {Bendinelli}, {Bonaccini},
  {Close}, {Currie}, \& {Parmeggiani}}]{starfinderpaper}
{Diolaiti}, E., {Bendinelli}, O., {Bonaccini}, D., {Close}, L., {Currie}, D.,
  \& {Parmeggiani}, G. 2000, \aaps, 147, 335

\bibitem[{{Draine}(1989)}]{Draine89}
{Draine}, B.~T. 1989, in ESA Special Publication, Vol. 290, Infrared
  Spectroscopy in Astronomy, ed. E.~{B{\"o}hm-Vitense}, 93--98

\bibitem[{{Dyson} \& {Williams}(1997)}]{Dyson-Williams}
{Dyson}, J.~E. \& {Williams}, D.~A. 1997, {The physics of the interstellar
  medium}, ed. {Dyson, J.~E.~\& Williams, D.~A.}

\bibitem[{{Egan} {et~al.}(2003){Egan}, {Price}, {Kraemer}, {Mizuno}, {Carey},
  {Wright}, {Engelke}, {Cohen}, \& {Gugliotti}}]{Egan03}
{Egan}, M.~P., {Price}, S.~D., {Kraemer}, K.~E., {Mizuno}, D.~R., {Carey},
  S.~J., {Wright}, C.~O., {Engelke}, C.~W., {Cohen}, M., \& {Gugliotti}, M.~G.
  2003, VizieR Online Data Catalog, 5114, 0

\bibitem[{{Ellingsen}(2006)}]{Ellingsen06}
{Ellingsen}, S.~P. 2006, \apj, 638, 241

\bibitem[{{Flaherty} {et~al.}(2007){Flaherty}, {Pipher}, {Megeath}, {Winston},
  {Gutermuth}, {Muzerolle}, {Allen}, \& {Fazio}}]{Flaherty07}
{Flaherty}, K.~M., {Pipher}, J.~L., {Megeath}, S.~T., {Winston}, E.~M.,
  {Gutermuth}, R.~A., {Muzerolle}, J., {Allen}, L.~E., \& {Fazio}, G.~G. 2007,
  \apj, 663, 1069

\bibitem[{{Foster} \& {Chevalier}(1993)}]{Foster-Chevalier93}
{Foster}, P.~N. \& {Chevalier}, R.~A. 1993, \apj, 416, 303

\bibitem[{{Froebrich} {et~al.}(2006){Froebrich}, {Schmeja}, {Smith}, \&
  {Klessen}}]{Froebrich06}
{Froebrich}, D., {Schmeja}, S., {Smith}, M.~D., \& {Klessen}, R.~S. 2006,
  \mnras, 368, 435

\bibitem[{{Harayama} {et~al.}(2008){Harayama}, {Eisenhauer}, \&
  {Martins}}]{Harayama08}
{Harayama}, Y., {Eisenhauer}, F., \& {Martins}, F. 2008, \apj, 675, 1319

\bibitem[{{Hoare} {et~al.}(2005){Hoare}, {Lumsden}, {Oudmaijer}, {Urquhart},
  {Busfield}, {Sheret}, {Clarke}, {Moore}, {Allsopp}, {Burton}, {Purcell},
  {Jiang}, \& {Wang}}]{Hoare05}
{Hoare}, M.~G., {Lumsden}, S.~L., {Oudmaijer}, R.~D., {Urquhart}, J.~S.,
  {Busfield}, A.~L., {Sheret}, T.~L., {Clarke}, A.~J., {Moore}, T.~J.~T.,
  {Allsopp}, J., {Burton}, M.~G., {Purcell}, C.~R., {Jiang}, Z., \& {Wang}, M.
  2005, in IAU Symposium, Vol. 227, Massive Star Birth: A Crossroads of
  Astrophysics, ed. {R.~Cesaroni, M.~Felli, E.~Churchwell, \& M.~Walmsley},
  370--375

\bibitem[{{Hosokawa} \& {Omukai}(2009)}]{H-O09}
{Hosokawa}, T. \& {Omukai}, K. 2009, \apj, 691, 823

\bibitem[{{Kahn}(1974)}]{Kahn74}
{Kahn}, F.~D. 1974, \aap, 37, 149

\bibitem[{{Kennicutt}(1998)}]{Kennicutt98}
{Kennicutt}, Jr., R.~C. 1998, \apj, 498, 541

\bibitem[{{Kothes} \& {Dougherty}(2007)}]{K-D07}
{Kothes}, R. \& {Dougherty}, S.~M. 2007, ArXiv e-prints, 0704.3073

\bibitem[{{Kroupa}(2001)}]{Kroupa01}
{Kroupa}, P. 2001, \mnras, 322, 231

\bibitem[{{Krumholz} {et~al.}(2009){Krumholz}, {Klein}, {McKee}, {Offner}, \&
  {Cunningham}}]{Krumholz09}
{Krumholz}, M.~R., {Klein}, R.~I., {McKee}, C.~F., {Offner}, S.~S.~R., \&
  {Cunningham}, A.~J. 2009, Science, 323, 754

\bibitem[{{Kuiper} {et~al.}(2011){Kuiper}, {Klahr}, {Beuther}, \&
  {Henning}}]{Kuiper11}
{Kuiper}, R., {Klahr}, H., {Beuther}, H., \& {Henning}, T. 2011, \apj, 732, 20

\bibitem[{{Kurtev} {et~al.}(2007){Kurtev}, {Borissova}, {Georgiev}, {Ortolani},
  \& {Ivanov}}]{Kurtev07}
{Kurtev}, R., {Borissova}, J., {Georgiev}, L., {Ortolani}, S., \& {Ivanov},
  V.~D. 2007, \aap, 475, 209

\bibitem[{{Lanz} \& {Hubeny}(2007)}]{L-H07}
{Lanz}, T. \& {Hubeny}, I. 2007, \apjs, 169, 83

\bibitem[{{Lumsden} {et~al.}(2002){Lumsden}, {Hoare}, {Oudmaijer}, \&
  {Richards}}]{Lumsden02}
{Lumsden}, S.~L., {Hoare}, M.~G., {Oudmaijer}, R.~D., \& {Richards}, D. 2002,
  \mnras, 336, 621

\bibitem[{{Lutz}(1999)}]{Lutz99}
{Lutz}, D. 1999, in ESA Special Publication, Vol. 427, The Universe as Seen by
  ISO, ed. P.~{Cox} \& M.~{Kessler}, 623--+

\bibitem[{{Martins} \& {Plez}(2006)}]{Martins-Plez06}
{Martins}, F. \& {Plez}, B. 2006, \aap, 457, 637

\bibitem[{{Martins} {et~al.}(2005){Martins}, {Schaerer}, \&
  {Hillier}}]{Martins05}
{Martins}, F., {Schaerer}, D., \& {Hillier}, D.~J. 2005, \aap, 436, 1049

\bibitem[{{McKee} \& {Tan}(2003)}]{M-T03}
{McKee}, C.~F. \& {Tan}, J.~C. 2003, \apj, 585, 850

\bibitem[{{Melena} {et~al.}(2008){Melena}, {Massey}, {Morrell}, \&
  {Zangari}}]{Melena08}
{Melena}, N.~W., {Massey}, P., {Morrell}, N.~I., \& {Zangari}, A.~M. 2008, \aj,
  135, 878

\bibitem[{{Messineo} {et~al.}(2009){Messineo}, {Davies}, {Ivanov}, {Figer},
  {Schuller}, {Habing}, {Menten}, \& {Petr-Gotzens}}]{Messineo09}
{Messineo}, M., {Davies}, B., {Ivanov}, V.~D., {Figer}, D.~F., {Schuller}, F.,
  {Habing}, H.~J., {Menten}, K.~M., \& {Petr-Gotzens}, M.~G. 2009, \apj, 697,
  701

\bibitem[{{Messineo} {et~al.}(2008){Messineo}, {Figer}, {Davies}, {Rich},
  {Valenti}, \& {Kudritzki}}]{Messineo08}
{Messineo}, M., {Figer}, D.~F., {Davies}, B., {Rich}, R.~M., {Valenti}, E., \&
  {Kudritzki}, R.~P. 2008, \apjl, 683, L155

\bibitem[{{Messineo} {et~al.}(2005){Messineo}, {Habing}, {Menten}, {Omont},
  {Sjouwerman}, \& {Bertoldi}}]{Messineo05}
{Messineo}, M., {Habing}, H.~J., {Menten}, K.~M., {Omont}, A., {Sjouwerman},
  L.~O., \& {Bertoldi}, F. 2005, \aap, 435, 575

\bibitem[{{Meynet} \& {Maeder}(2000)}]{Mey-Mae00}
{Meynet}, G. \& {Maeder}, A. 2000, \aap, 361, 101

\bibitem[{{Motoyama} \& {Yoshida}(2003)}]{Motoyama-Yoshida03}
{Motoyama}, K. \& {Yoshida}, T. 2003, \mnras, 344, 461

\bibitem[{{Mottram} {et~al.}(2011{\natexlab{a}}){Mottram}, {Hoare}, {Davies},
  {Lumsden}, {Oudmaijer}, {Urquhart}, {Moore}, {Cooper}, \&
  {Stead}}]{Mottram11L}
{Mottram}, J.~C., {Hoare}, M.~G., {Davies}, B., {Lumsden}, S.~L., {Oudmaijer},
  R.~D., {Urquhart}, J.~S., {Moore}, T.~J.~T., {Cooper}, H.~D.~B., \& {Stead},
  J.~J. 2011{\natexlab{a}}, \apjl, 730, L33+

\bibitem[{{Mottram} {et~al.}(2007){Mottram}, {Hoare}, {Lumsden}, {Oudmaijer},
  {Urquhart}, {Sheret}, {Clarke}, \& {Allsopp}}]{Mottram07}
{Mottram}, J.~C., {Hoare}, M.~G., {Lumsden}, S.~L., {Oudmaijer}, R.~D.,
  {Urquhart}, J.~S., {Sheret}, T.~L., {Clarke}, A.~J., \& {Allsopp}, J. 2007,
  \aap, 476, 1019

\bibitem[{{Mottram} {et~al.}(2011{\natexlab{b}}){Mottram}, {Hoare}, {Urquhart},
  {Lumsden}, {Oudmaijer}, {Robitaille}, {Moore}, {Davies}, \&
  {Stead}}]{Mottram11}
{Mottram}, J.~C., {Hoare}, M.~G., {Urquhart}, J.~S., {Lumsden}, S.~L.,
  {Oudmaijer}, R.~D., {Robitaille}, T.~P., {Moore}, T.~J.~T., {Davies}, B., \&
  {Stead}, J. 2011{\natexlab{b}}, \aap, 525, A149+

\bibitem[{{Plume} {et~al.}(1997){Plume}, {Jaffe}, {Evans}, {Martin-Pintado}, \&
  {Gomez-Gonzalez}}]{Plume97}
{Plume}, R., {Jaffe}, D.~T., {Evans}, II, N.~J., {Martin-Pintado}, J., \&
  {Gomez-Gonzalez}, J. 1997, \apj, 476, 730

\bibitem[{{Qiu} {et~al.}(2011){Qiu}, {Zhang}, \& {Menten}}]{Qiu11}
{Qiu}, K., {Zhang}, Q., \& {Menten}, K.~M. 2011, \apj, 728, 6

\bibitem[{{Rathborne} {et~al.}(2010){Rathborne}, {Jackson}, {Chambers},
  {Stojimirovic}, {Simon}, {Shipman}, \& {Frieswijk}}]{Rathborne10}
{Rathborne}, J.~M., {Jackson}, J.~M., {Chambers}, E.~T., {Stojimirovic}, I.,
  {Simon}, R., {Shipman}, R., \& {Frieswijk}, W. 2010, \apj, 715, 310

\bibitem[{{Reid} {et~al.}(2009){Reid}, {Menten}, {Zheng}, {Brunthaler},
  {Moscadelli}, {Xu}, {Zhang}, {Sato}, {Honma}, {Hirota}, {Hachisuka}, {Choi},
  {Moellenbrock}, \& {Bartkiewicz}}]{Reid09}
{Reid}, M.~J., {Menten}, K.~M., {Zheng}, X.~W., {Brunthaler}, A., {Moscadelli},
  L., {Xu}, Y., {Zhang}, B., {Sato}, M., {Honma}, M., {Hirota}, T.,
  {Hachisuka}, K., {Choi}, Y.~K., {Moellenbrock}, G.~A., \& {Bartkiewicz}, A.
  2009, \apj, 700, 137

\bibitem[{{Robitaille} \& {Whitney}(2010)}]{R-W10}
{Robitaille}, T.~P. \& {Whitney}, B.~A. 2010, \apjl, 710, L11

\bibitem[{{Robitaille} {et~al.}(2006){Robitaille}, {Whitney}, {Indebetouw},
  {Wood}, \& {Denzmore}}]{Robitaille06}
{Robitaille}, T.~P., {Whitney}, B.~A., {Indebetouw}, R., {Wood}, K., \&
  {Denzmore}, P. 2006, \apjs, 167, 256

\bibitem[{{Schaller} {et~al.}(1992){Schaller}, {Schaerer}, {Meynet}, \&
  {Maeder}}]{Schaller92}
{Schaller}, G., {Schaerer}, D., {Meynet}, G., \& {Maeder}, A. 1992, \aaps, 96,
  269

\bibitem[{{Schmeja} \& {Klessen}(2004)}]{S-K04}
{Schmeja}, S. \& {Klessen}, R.~S. 2004, \aap, 419, 405

\bibitem[{{Shu}(1977)}]{Shu77}
{Shu}, F.~H. 1977, \apj, 214, 488

\bibitem[{{Stahler} {et~al.}(2000){Stahler}, {Palla}, \& {Ho}}]{Stahler00}
{Stahler}, S.~W., {Palla}, F., \& {Ho}, P.~T.~P. 2000, Protostars and Planets
  IV, 327

\bibitem[{{Tapia} {et~al.}(2003){Tapia}, {Roth}, {V{\'a}zquez}, \&
  {Feinstein}}]{Tapia03}
{Tapia}, M., {Roth}, M., {V{\'a}zquez}, R.~A., \& {Feinstein}, A. 2003, \mnras,
  339, 44

\bibitem[{{Taylor} \& {Cordes}(1993)}]{T-C93}
{Taylor}, J.~H. \& {Cordes}, J.~M. 1993, \apj, 411, 674

\bibitem[{{Urquhart} {et~al.}(2007{\natexlab{a}}){Urquhart}, {Busfield},
  {Hoare}, {Lumsden}, {Clarke}, {Moore}, {Mottram}, \&
  {Oudmaijer}}]{Urquhart07a}
{Urquhart}, J.~S., {Busfield}, A.~L., {Hoare}, M.~G., {Lumsden}, S.~L.,
  {Clarke}, A.~J., {Moore}, T.~J.~T., {Mottram}, J.~C., \& {Oudmaijer}, R.~D.
  2007{\natexlab{a}}, \aap, 461, 11

\bibitem[{{Urquhart} {et~al.}(2007{\natexlab{b}}){Urquhart}, {Busfield},
  {Hoare}, {Lumsden}, {Oudmaijer}, {Moore}, {Gibb}, {Purcell}, {Burton}, \&
  {Marechal}}]{Urquhart07b}
{Urquhart}, J.~S., {Busfield}, A.~L., {Hoare}, M.~G., {Lumsden}, S.~L.,
  {Oudmaijer}, R.~D., {Moore}, T.~J.~T., {Gibb}, A.~G., {Purcell}, C.~R.,
  {Burton}, M.~G., \& {Marechal}, L.~J.~L. 2007{\natexlab{b}}, \aap, 474, 891

\bibitem[{{Urquhart} {et~al.}(2008){Urquhart}, {Busfield}, {Hoare}, {Lumsden},
  {Oudmaijer}, {Moore}, {Gibb}, {Purcell}, {Burton}, {Mar{\'e}chal}, {Jiang},
  \& {Wang}}]{Urquhart08}
{Urquhart}, J.~S., {Busfield}, A.~L., {Hoare}, M.~G., {Lumsden}, S.~L.,
  {Oudmaijer}, R.~D., {Moore}, T.~J.~T., {Gibb}, A.~G., {Purcell}, C.~R.,
  {Burton}, M.~G., {Mar{\'e}chal}, L.~J.~L., {Jiang}, Z., \& {Wang}, M. 2008,
  \aap, 487, 253

\bibitem[{{Urquhart} {et~al.}(2009{\natexlab{a}}){Urquhart}, {Hoare},
  {Lumsden}, {Oudmaijer}, {Moore}, {Brook}, {Mottram}, {Davies}, \&
  {Stead}}]{Urquhart09b}
{Urquhart}, J.~S., {Hoare}, M.~G., {Lumsden}, S.~L., {Oudmaijer}, R.~D.,
  {Moore}, T.~J.~T., {Brook}, P.~R., {Mottram}, J.~C., {Davies}, B., \&
  {Stead}, J.~J. 2009{\natexlab{a}}, \aap, 507, 795

\bibitem[{{Urquhart} {et~al.}(2009{\natexlab{b}}){Urquhart}, {Hoare},
  {Purcell}, {Lumsden}, {Oudmaijer}, {Moore}, {Busfield}, {Mottram}, \&
  {Davies}}]{Urquhart09a}
{Urquhart}, J.~S., {Hoare}, M.~G., {Purcell}, C.~R., {Lumsden}, S.~L.,
  {Oudmaijer}, R.~D., {Moore}, T.~J.~T., {Busfield}, A.~L., {Mottram}, J.~C.,
  \& {Davies}, B. 2009{\natexlab{b}}, \aap, 501, 539

\bibitem[{{Whitney} {et~al.}(2003){Whitney}, {Wood}, {Bjorkman}, \&
  {Wolff}}]{Whitney03}
{Whitney}, B.~A., {Wood}, K., {Bjorkman}, J.~E., \& {Wolff}, M.~J. 2003, \apj,
  591, 1049

\bibitem[{{Whitworth} \& {Ward-Thompson}(2001)}]{Whitworth-WardThompson01}
{Whitworth}, A.~P. \& {Ward-Thompson}, D. 2001, \apj, 547, 317

\bibitem[{{Wolfire} \& {Cassinelli}(1987)}]{W-C87}
{Wolfire}, M.~G. \& {Cassinelli}, J.~P. 1987, \apj, 319, 850

\bibitem[{{Wood} \& {Churchwell}(1989)}]{W-C89}
{Wood}, D.~O.~S. \& {Churchwell}, E. 1989, \apjs, 69, 831

\bibitem[{{Zinnecker} \& {Yorke}(2007)}]{Z-Y07}
{Zinnecker}, H. \& {Yorke}, H.~W. 2007, \araa, 45, 481

\end{thebibliography}

\appendix

\section[]{Computation of line-of-sight extinction} \label{sec:appext}
In calculating the extinction towards each object in our simulation,
we cannot simply use 2-D extinction maps of the Galactic Plane, as
these do not give any indication of how extinction varies with
distance along a given sightline. Instead, we use test points with known
distances and extinctions, and derive a relation between the
intervening gas column density and the line of sight extinction by
assuming that the extinction is some function of the interstellar gas
column density between the Earth and the object. This relation should
be linear, as they are both directly proportional to the optical
depth. 

\begin{figure}
  \centering
  \includegraphics[width=8.5cm]{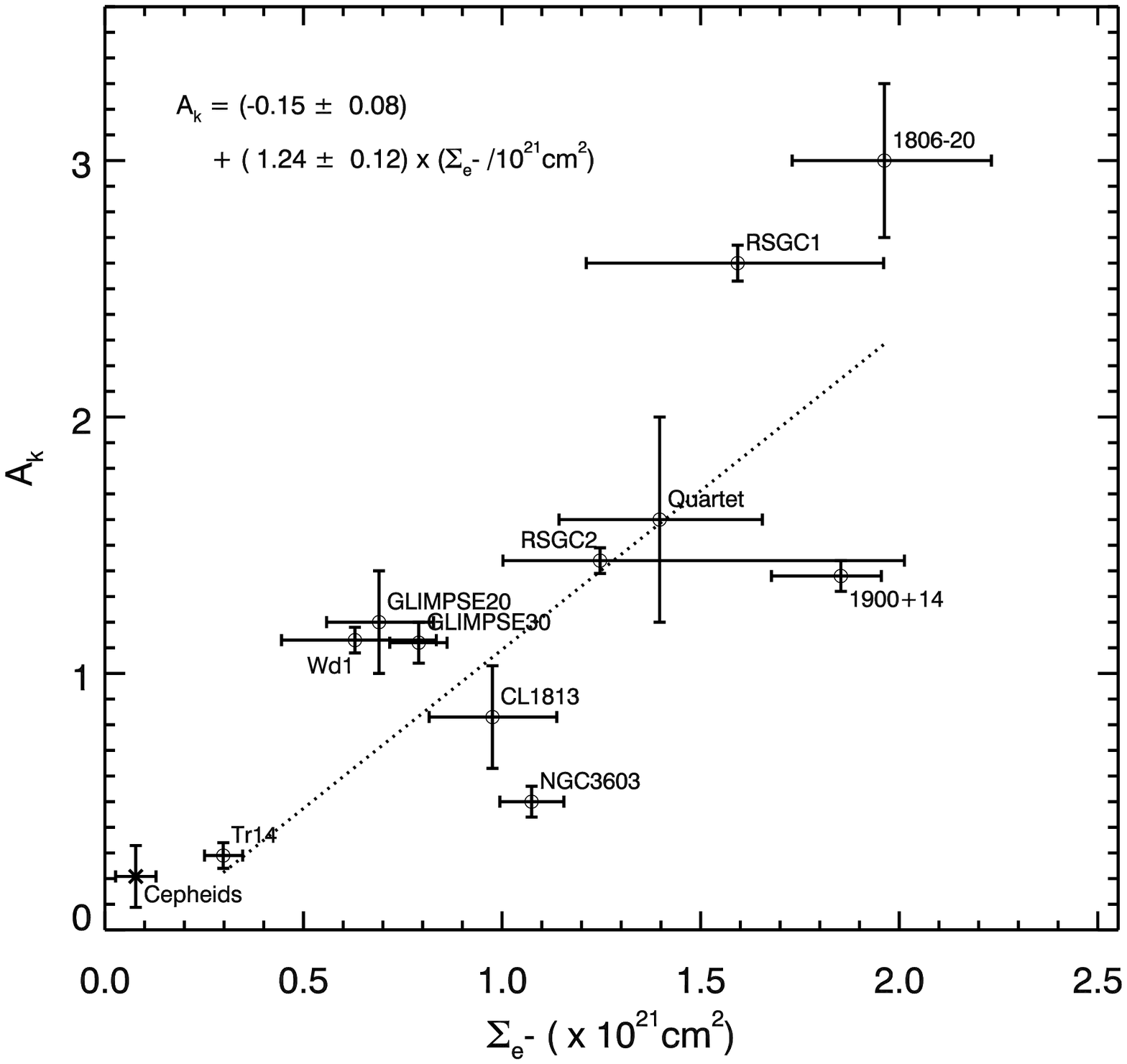}
  \caption{The $K$-band extinction of a sample of Galactic clusters
    versus the inferred gas column density from our model of the
    Galactic gas distribution. }
  \label{fig:compext}
\end{figure}

\begin{table}
  \begin{tabular}{lccccc}
    \hline\hline
    Name & $l$ (\degr) & $b$ (\degr) & D$_{\odot}$(kpc)        & A$_{k}$ & Ref \\
    \hline 
    1806-20	 & 10.00  & -0.24 &  8.7$^{+1.8}_{-1.5}$ & 3.0$\pm$0.3   & 1 \\
    CL1813	 & 12.77  & -0.01 &  4.7$\pm$0.4       & 0.83$\pm$0.2  & 2 \\
    Quartet	 & 24.90  & +0.12 &  6.1$\pm$0.6       & 1.6$\pm$0.4   & 3  \\	
    RSGC1	 & 25.27  & -0.16 &  6.6$\pm$0.9       & 2.60$\pm$0.07 & 4 \\	
    RSGC2	 & 26.19  & -0.07 &  5.8$^{+1.8}_{-0.6}$ & 1.44$\pm$0.05 & 5 \\ 
    1900+14	 & 43.02  & +0.77 & 12.5$\pm$1.7       & 1.38$\pm$0.06 & 6 \\
    GLIMPSE20    & 44.16  & -0.07 &  4.5$\pm$0.7       & 1.2$\pm$0.2   & 3 \\
    Tr14	 & 287.40 & -0.58 &  2.7$\pm$0.3       & 0.29$\pm$0.05 & 7,8 \\	
    NGC3603	 & 291.63 & -0.53 &  7.6$\pm$0.5       & 0.5$\pm$0.06  & 9,10 \\ 
    GLIMPSE30    & 298.76 & -0.41 &  7.2$\pm$0.9       & 1.12$\pm$0.08 & 11 \\
    Wd1    	 & 339.55 & -0.40 &  3.9$\pm$0.7       & 1.13$\pm$0.05 & 12,13 \\
    \hline
  \end{tabular}
  \caption{Compilation of young Galactic clusters used in the
    calibration of free-electron column density with K-band
    extinction. References are, 1: \citet{Bibby08}; 2:
    \citet{Messineo08}; 3: \citet{Messineo09}; 4: \citet{RSGC2paper};
    5: \citet{RSGC1paper}; 6: \citet{SGR1900paper}; 7:
    \citet{Ascenso07}; 8: \citet{Tapia03}; 9: \citet{Melena08}; 10:
    \citet{Harayama08}; 11: \citet{Kurtev07}; 12: \citet{Brandner08};
    13: \citet{K-D07}} 
  \label{tab:clust}
\end{table}

\begin{figure*}
  \centering
  \includegraphics[width=17cm]{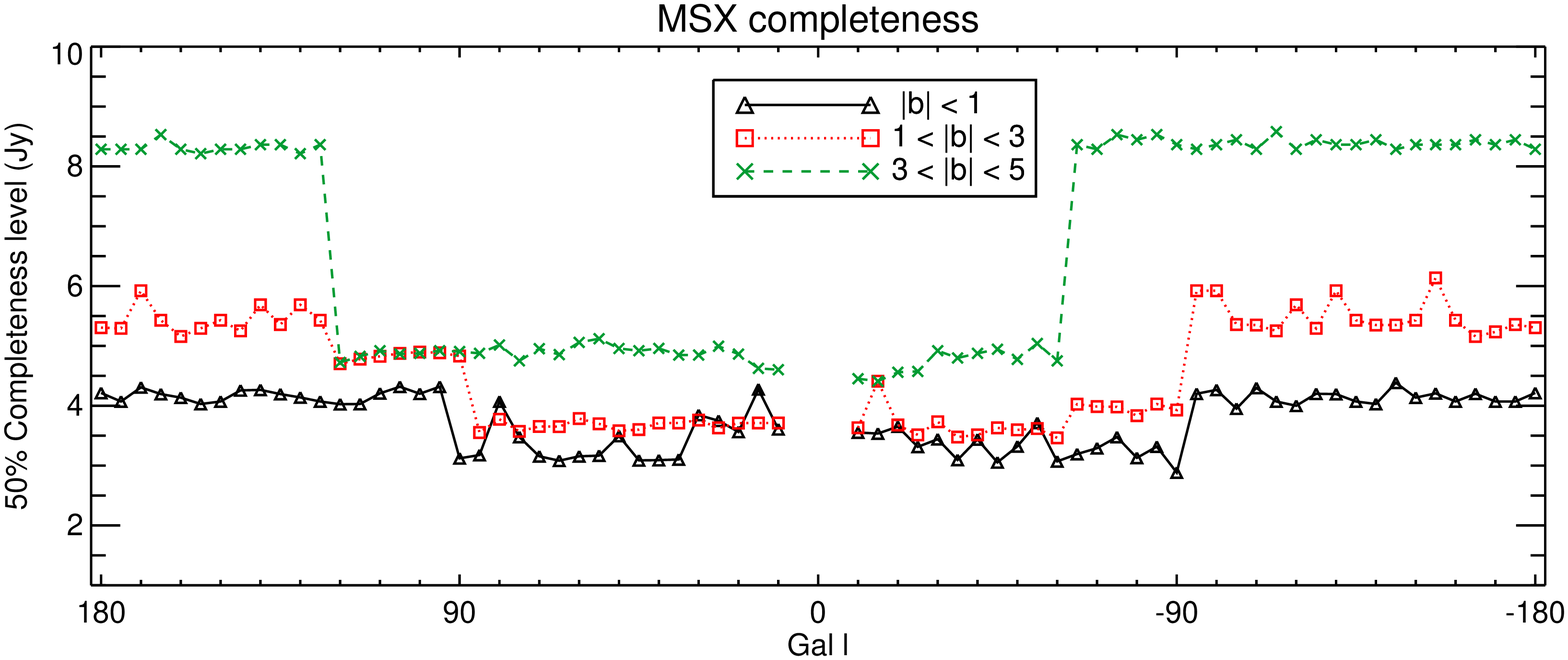}
  \caption{Illustration of the 50\% completeness level of the
    \MSX\ survey at 21\um, as a function of both $l$ and $b$. }
  \label{fig:msxcomp}
\end{figure*}

For the test points, we must use objects that span a large range of
distances and extinctions, as we will need to use this function to
compute the extinctions to objects at the far side of the Galaxy. For
this reason, we have chosen to use young stellar clusters as the test
points. They are very bright in the near-IR, and so can be observed at
very large extinctions ($A_{V} \sim 30$); while spectroscopic
observations of the stellar content allow the determination of both
spectrophotometric and kinematic distances. 

In Table \ref{tab:clust} we list a sample of clusters with known
distances and $K$-band extinctions which we use to derive the relation
between extinction and column density. For each cluster, we use the
model of the Galactic electron density described in
Sect.\ \ref{sec:galcube} to calculate the integrated gas column
density along the line-of-sight from Earth to the cluster,
$\Sigma_{e^{-}}$. We also complement these data with that from
cepheids, using the data from \citet{Berdnikov00}.

In Fig.\ \ref{fig:compext} we plot $A_{K}$ against \sige\ for each
cluster in the sample. For the cepheids, which are all at low column
densities, we take the average of the whole sample. As expected, the
relation is consistent with being linear, where we find,

\begin{equation}
A_{K} = (-0.15 \pm 0.08) + (1.24 \pm 0.12) \times
(\Sigma_{e^{-}}/10^{21} {\rm cm}^{2})
\end{equation}

\noindent Note that if the gas distribution is altered this relation
must be recalculated. To convert the $K$-band extinction $A_{K}$ to
the extinction at 21\microns\ $A_{21}$, we must know the ratio
$A_{21}$/$A_{K}$ for interstellar extinction. Though not well studied,
there seems to be some agreement in the literature that this value is
$\sim$0.4-0.6 from studies of evolved stars, star-forming regions, the
Galactic Centre, and the diffuse interstellar medium
\citep{Draine89,Lutz99,Flaherty07,Chapman09,Messineo05}. For the
remainder of this work, we assume $A_{21}$/$A_{K} = 0.5$.

\section[]{Completeness of the \MSX\ survey at 21\um} \label{sec:msx}
In order to accurately predict whether or not a source would be picked
up by the \MSX\ survey, we must first know the detection limits of the
survey. The survey is not isotropic, as some regions of the sky were
scanned more times than others, so we must determine the completeness as a
function of viewing angle. 

To measure the completeness of the \MSX\ survey, we took
representitive cutout images 1\degr$\times$1\degr\ in size. We then
analyzed the images with the photometry package {\sc starfinder}
\citep{starfinderpaper}. The input parameters were adjusted until the
closest match was achieved between the {\sc starfinder} and
\MSX\ source lists. The typical overlap between the sources was
$\ga$80\%.

Using a point-spread-function (PSF) measured from the test image, we
then inserted fake stars into the image of a set brightness. No more
than 40 stars were added at one time so as not to affect the
crowding. We then ran the {\sc starfinder} algorithm on the image and
measured the recovery rate. The process was repeated five times per
fake-star brightness interval, and six different brightness levels
were tested between 12Jy and 2Jy. This process was repeated on a large
number of fields, spaced at intervals of 5\degr\ in $l$ and 2\degr\ in
$b$, a total of 345 images in all.

\end{document}